\documentclass[twocolumn, twocolappendix]{aastex631}

\newcommand{\oiii}{[O\,\textsc{iii}]}
\newcommand{\oi}{[O\,\textsc{i}]}
\newcommand{\nii}{[N\,\textsc{ii}]}
\newcommand{\sii}{[S\,\textsc{ii}]}
\newcommand{\ha}{H$\alpha$}
\newcommand{\hb}{H$\beta$}
\newcommand{\hg}{H$\gamma$}

\newcommand{\cii}{[C\,\textsc{ii}]}
\newcommand{\feii}{Fe\,\textsc{ii}}
\newcommand{\heii}{He\,\textsc{ii}}
\newcommand{\sersic}{S\'ersic}
\newcommand{\targ}{GNz7q}

\newcommand{\galfits}{\textsc{GalfitS}}
\newcommand{\cigale}{\texttt{cigale}}
\usepackage{hyperref}

\shorttitle{Red Quasar}
\shortauthors{Fei et al.}
\graphicspath{{./}{figures/}}

\usepackage{amsmath}

\begin{document}

\title{
Direct pathway to the Early Supermassive Black Holes: \\ 
A Red Super-Eddington Quasar in a Massive Starburst Host at \boldmath $z=7.2$ 
}
\author[0000-0001-7232-5355]{Qinyue Fei}
\affiliation{David A. Dunlap Department of Astronomy and Astrophysics, University of Toronto, 50 St. George Street, Toronto, Ontario, M5S 3H4, Canada}
\email{qyfei.astro@gmail.com}


\author[0000-0001-7201-5066]{Seiji Fujimoto}
\affiliation{David A. Dunlap Department of Astronomy and Astrophysics, University of Toronto, 50 St. George Street, Toronto, Ontario, M5S 3H4, Canada}
\affiliation{Dunlap Institute for Astronomy and Astrophysics, 50 St. George Street, Toronto, Ontario, M5S 3H4, Canada}

\author[0000-0003-2680-005X]{Gabriel Brammer}
\affiliation{Cosmic Dawn Center (DAWN), Niels Bohr Institute, University of Copenhagen, Jagtvej 128, K{\o}benhavn N, DK-2200, Denmark}

\author[0000-0001-8496-4162]{Ruancun Li}
\affiliation{Max-Planck-Institut f{\"u}r extraterrestrische Physik, Gie{\ss}enbachstra{\ss}e 1, 85748 Garching bei M{\"u}nchen, Germany}

\author[0000-0001-6947-5846]{Luis C. Ho}
\affiliation{Kavli Institute for Astronomy and Astrophysics, Peking University, Beijing 100871, China}
\affiliation{Department of Astronomy, School of Physics, Peking University, Beijing 100871, China}

\author[0000-0003-0212-2979]{Volker Bromm}
\affiliation{Department of Astronomy, University of Texas, Austin, TX 78712, USA}
\affiliation{Weinberg Institute for Theoretical Physics, University of Texas, Austin, TX 78712, USA}
\affiliation{Cosmic Frontier Center, The University of Texas at Austin, Austin, TX 78712, USA}

\author[0000-0002-7093-1877]{Javier \'{A}lvarez-M\'{a}rquez}
\affiliation{Centro de Astrobiolog\'{\i}a (CAB), CSIC-INTA, Ctra. de Ajalvir km 4, Torrej\'{o}n de Ardoz, E-28850, Madrid, Spain}

\author[0000-0003-3983-5438]{Yoshihisa Asada}
\affiliation{David A. Dunlap Department of Astronomy and Astrophysics, University of Toronto, 50 St. George Street, Toronto, Ontario, M5S 3H4, Canada}
\affiliation{Dunlap Institute for Astronomy and Astrophysics, 50 St. George Street, Toronto, Ontario, M5S 3H4, Canada}

\author[0000-0001-6813-875X]{Guillermo Barro}
\affiliation{University of the Pacific, Stockton, CA 90340 USA}

\author[0000-0002-9090-4227]{Luis Colina}
\affiliation{Centro de Astrobiolog\'{\i}a (CAB), CSIC-INTA, Ctra. de Ajalvir km 4, Torrej\'{o}n de Ardoz, E-28850, Madrid, Spain}

\author[0000-0001-8460-1564]{Pratika Dayal}
\affiliation{Canadian Institute for Theoretical Astrophysics, 60 St George St, University of Toronto, Toronto, ON M5S 3H8, Canada}
\affiliation{David A. Dunlap Department of Astronomy and Astrophysics, University of Toronto, 50 St George St, Toronto ON M5S 3H8, Canada }
\affiliation{Department of Physics, 60 St George St, University of Toronto, Toronto, ON M5S 3H8, Canada}

\author[0000-0001-8519-1130]{Steven L. Finkelstein}
\affiliation{Department of Astronomy, The University of Texas at Austin, Austin, TX, USA}
\affiliation{Cosmic Frontier Center, The University of Texas at Austin, Austin, TX, USA}

\author[0000-0002-8149-8298]{Johan~P.~U.~Fynbo}
\affiliation{Cosmic DAWN Center}
\affiliation{Niels Bohr Institute, University of Copenhagen, Jagtvej 128, 2200 Copenhagen N, Denmark}

\author[0000-0002-9122-1700]{Michele Ginolfi}
\affiliation{Universit\'{'a} di Firenze, Dipartimento di Fisica e Astronomia, via G. Sansone 1, 50019 Sesto Fiorentino, Florence, Italy}
\affiliation{INAF -- Arcetri Astrophysical Observatory, Largo E. Fermi 5, I-50125, Florence, Italy}

\author[0000-0000-0000-0001]{Kohei Inayoshi}
\affiliation{Kavli Institute for Astronomy and Astrophysics, Peking University, Beijing 100871, China}

\author[0000-0002-5588-9156]{Vasily Kokorev}
\affiliation{Department of Astronomy, University of Texas, Austin, TX 78712, USA}
\affiliation{Cosmic Frontier Center, The University of Texas at Austin, Austin, TX 78712, USA}

\author[0000-0002-9393-6507]{Gene C. K. Leung}
\affiliation{Department of Astronomy, The University of Texas at Austin, Austin, TX 78712, USA}
\affiliation{MIT Kavli Institute for Astrophysics and Space Research, 77 Massachusetts Ave., Cambridge, MA 02139, USA}

\author[0000-0003-2871-127X]{Jorryt Matthee} 
\affiliation{Institute of Science and Technology Austria (ISTA), Am Campus 1, 3400 Klosterneuburg, Austria}

\author[0000-0001-5492-4522]{Romain A. Meyer}
\affiliation{Department of Astronomy, University of Geneva, Chemin Pegasi 51, 1290 Versoix, Switzerland}

\author[0000-0003-3997-5705]{Rohan P.~Naidu}\altaffiliation{Hubble Fellow}
\affiliation{
MIT Kavli Institute for Astrophysics and Space Research, 70 Vassar Street, Cambridge, MA 02139, USA}

\author[0000-0003-2984-6803]{Masafusa Onoue}
\affiliation{Waseda Institute for Advanced Study (WIAS), Waseda University, 1-21-1, Nishi-Waseda, Shinjuku, Tokyo 169-0051, Japan}
\affiliation{Kavli Institute for the Physics and Mathematics of the Universe (WPI), The University of Tokyo, Institutes for Advanced Study, The University of Tokyo, Kashiwa, Chiba 277-8583, Japan}
\affiliation{Kavli Institute for Astronomy and Astrophysics, Peking University, Beijing 100871, China}
\affiliation{Center for Data-Driven Discovery, Kavli IPMU (WPI), UTIAS, The University of Tokyo, Kashiwa, Chiba 277-8583, Japan}

\author[0000-0003-4528-5639]{Pablo G. P\'erez-Gonz\'alez}
\affiliation{Centro de Astrobiolog\'{\i}a (CAB), CSIC-INTA, Ctra. de Ajalvir km 4, Torrej\'{o}n de Ardoz, E-28850, Madrid, Spain}

\author[0000-0003-3780-6801]{Charles L. Steinhardt}
\affiliation{Department of Physics and Astronomy, University of Missouri, 701 S. College Ave., Columbia, MO 65201, USA}

\author[0000-0001-6477-4011]{Francesco Valentino }
\affiliation{Cosmic Dawn Center}
\affiliation{DTU Space, Technical University of Denmark, Elektrovej 327, DK2800 Kgs. Lyngby, Denmark}

\author[0000-0003-4793-7880]{Fabian Walter}
\affiliation{Max Planck Institut f\"ur Astronomie, K\"onigstuhl 17, D-69117 Heidelberg, Germany}
\affiliation{California Institute of Technology, Pasadena, CA 91125, USA}
\affiliation{National Radio Astronomy Observatory, Pete V. Domenici Array Science Center, P.O. Box O, Socorro, NM 87801, USA}

\author[0000-0003-1207-5344]{Mengyuan Xiao}
\affiliation{Department of Astronomy, University of Geneva, Chemin Pegasi 51, 1290 Versoix, Switzerland}

\author[0000-0002-4321-3538]{Haowen Zhang}
\affiliation{Canadian Institute for Theoretical Astrophysics, 60 St George St, University of Toronto, Toronto, ON M5S 3H8, Canada}

\begin{abstract}
We present a panchromatic optical--mm characterization of \targ, a recently identified X-ray weak, rapidly growing red quasar embedded within a dusty starburst galaxy at $z=7.1899$, using the full suite of JWST/NIRCam, NIRSpec, MIRI, and archival NOEMA observations.
Our deep NIRSpec/G395M spectroscopy reveals unambiguous broad Balmer emission (FWHM $=2221\pm20$~km~s$^{-1}$), confirming a super-Eddington accreting black hole ($\lambda_{\rm Edd}=2.7\pm0.4$) with a mass of $\log(M_{\rm BH}/M_{\odot})=7.55\pm0.34$, using accretion-rate corrected BH mass estimators. 
After subtracting the point source, we robustly detect stellar emission from the host galaxy across multiple NIRCam and MIRI filters. 
Our joint morphological–spectral analysis yields a stellar mass of $\log(M_{\ast}/M_\odot)=10.5\pm0.4$ and an intense star formation rate of ${\rm SFR}=330\pm97~M_\odot~\mathrm{yr^{-1}}$, confirming the host as a massive, dusty starburst galaxy. 
We find that \targ\ lies on the local $M_{\rm BH}$–$M_\ast$ relation ($M_{\rm BH}/M_\ast \simeq 0.001$) and is well positioned to evolve into the locus of massive SDSS quasars with $\log(M_{\rm BH}/M_\odot)\approx9$ and $M_\ast\approx10^{11}\,M_\odot$ at $z\sim6$, owing to its remarkably rapid growth in both the black hole and its host galaxy.
This stands in stark contrast to many recently reported JWST AGN populations at similar redshifts, including the little red dots (LRDs), whose weak or undetected star formation makes it difficult for them to grow into the massive galaxies hosting SDSS-like quasars.
These results suggest that \targ\ marks a rare, pivotal phase of early BH–galaxy co-evolution, plausibly providing a crucial direct pathway to the supermassive black hole systems within the first billion years of the Universe. 
\end{abstract}


\section{Introduction}
\label{sec1: intro}


The discovery of luminous quasars at redshift $z \gtrsim 6$ has fundamentally challenged our understanding of the early Universe \citep[e.g.,][]{Smith2019, Woods2019}. The existence of supermassive black holes (SMBHs) with masses exceeding $10^9\, M_\odot$ less than 1~Gyr after the Big Bang \citep[e.g.,][]{Fan+2001, Fan+2006, Mortlock+2011, Banados+2018, Wang+2021} necessitates specific mechanisms for rapid black hole growth. 
Theoretical models suggest that these giants grew from either heavy seeds, arising from the direct collapse of massive primordial gas clouds, or light seeds originating in stellar remnants that underwent continuous, super-Eddington accretion \citep[e.g.,][]{Volonteri2012, Inayoshi+2020}. 
In addition, these high-$z$ quasars appear to host `overmassive' BHs \citep{Yue+2024_QSO, Stone+2024_QSO, Li+2025_galfits}, relative to the stellar mass of their host galaxies when compared to the local $M_{\rm BH}-M_*$ relations \citep{Kormendy+2013, Reines+2015, Greene+2020, Zhuang+2023_AGN}. 

The launch of JWST has opened a new window into this field, unveiling a previously hidden population of fainter AGNs at high redshifts \citep[e.g., ][]{Harikane+2023, Larson+2023, Kokorev+2023, Kocevski+2023, Furtak+2024_len_LRD, Maiolino+2024_BH, Maiolino+2024, Matthee+2024_LRD, Lin+2024_BHE, Lin+2025_BHE3D, Taylor+2025_AGN, Taylor+2025_HzLRD, Hviding+2025_LRD, Fei+2025}. 
Identified by broad rest-frame optical Balmer lines, these sources enable the study of lower-mass black holes ($M_{\rm BH} \sim 10^5 - 10^8 M_\odot$) in the early Universe. 
In contrast to massive, luminous quasars that display mature features \citep{Bosman+2024_qsoBLR, Bosman+2025_qso}, these objects likely represent a nascent phase of black hole growth \citep{Inayoshi2025_LRD, Jeon2025}.

Many of these lower-luminosity AGNs reside in low-mass galaxies ($\sim10^8\,M_\odot$), revealing BH-to-stellar mass ratios ($M_{\rm BH}/M_*$) that deviate more dramatically from the local relationship, with some objects exhibiting values 10 to 100 times higher than the local benchmark \citep[e.g., ][]{Izumi+2019_Mdyn, Maiolino+2024_BH, Chen+2025_lrd, LiJ+2025b}, though this deviation may be less extreme if BH masses are systematically lower by $\sim1$ dex as recently suggested \citep[e.g., ][]{Greene+2025_lrd_lbol, Umeda+2025_lrd, Asada+2026_lrd}. 
The presence of such high $M_{\rm BH}/M_*$ ratios may directly support the existence of `heavy seeds' formed via Direct Collapse Black Holes (DCBHs), providing a possible mechanism for the rapid early growth of SMBHs \citep{Pacucci+2024, Maiolino+2024_BH, Jeon2025_DCBH}, or implicate more exotic formation channels \citep[e.g., the core collapse of self-interacting dark matter;][]{Jiang+2025_ccSIDM} or even primordial BHs \citep[e.g., ][]{Bean+2002_pBH, Carr+2020_pBH, Dayal2024_pBH, Zhang2025_PBH}. 
However, we need to consider that this observed deviation could be partially driven by observational biases, since flux-limited surveys may preferentially detect systems where the central engine outshines the host, artificially skewing samples toward high mass ratios \citep{LiJ+2025b, Silverman+2025}.

Crucially, whether these faint, overmassive sources represent the direct progenitors of the luminous quasars observed at $z\sim6$ remains an open question. 
To fully bridge these populations, it is essential to investigate the co-evolution of both the BH and the host galaxy, yet significant physical discrepancies exist between the two regimes. 
Unlike the vigorous starbursts associated with luminous quasar host galaxies \citep[e.g., ][]{Walter+2004_QSO, Wang+2010_qso, Wang+2013_qso, Decarli+2018_qso, Venemans+2020_qso, Neeleman+2021_qso_kin, Fei+2025_alma}, JWST-discovered AGNs often exhibit generally low star formation activity, characterized by strong upper limits on their rest-frame IR-derived SFRs \citep{Akins+2025, Casey+2025}. 
Furthermore, these sources usually display low Eddington ratios ($\lambda_{\rm Edd} \approx 0.1$; \citealt{Harikane+2023, Maiolino+2024}), contrasting with the rapid accretion required to build SMBHs. 

This ambiguity is further compounded by systematic uncertainties in deriving BH properties at these epochs. 
Standard estimators rely on empirical scaling relations calibrated in the local Universe, which may be inapplicable to high-$z$ faint AGNs given their distinct spectral characteristics\footnote{Direct BH mass measurements via resolved JWST integral-field kinematics can provide crucial cross-calibration and validation \citep{Juodzbalis2025_Direct}.}, such as deviant broad emission-line profiles \citep{Rusakov+2025_LRD, Kokorev+2025_BH*} and complex absorption features \citep{Wang+2025_LRD, D'Eugenio+2025_LRD_FeII}. 
Theoretical frameworks describing these systems introduce additional complexity \citep{Inayoshi+2025_LRD, Naidu+2025_BH*, Chang+2025_LRD}; these model dependencies imply uncertainties in BH mass and accretion rates ($\dot{\mathcal{M}}$) of up to $\sim 1$ dex \citep{Greene+2025_lrd_lbol}. 
Consequently, to bridge this gap and empirically verify the transition from faint seeds to luminous giants, it is crucial to investigate rare `transitional' sources \citep{Fu+2025_lrd}, which occupy the intermediate parameter space and offer simultaneous, robust constraints on both the central AGN and the host galaxy.
Comparable systematic uncertainties affect the measurement of the host galaxies. 
The decomposition of the host galaxy is inherently degenerate due to the overwhelming AGN continuum and limitations in Point Spread Function (PSF) subtraction \citep{Mechtley+2012_qso, Marshall+2020_qso}. 
Furthermore, these photometric ambiguities introduce significant scatter in the physical properties derived via Spectral Energy Distribution (SED) fitting \citep{Ding+2023_nat, Ding+2025_shellq}.

In addition to these systematic uncertainties in the measurements, the characterization of the high-redshift AGNs is currently limited by the survey volume of JWST. 
While the unprecedented sensitivity of JWST has revolutionized our ability to probe faint, low-mass galaxies ($M_* \lesssim 10^9\, M_\odot$) and low-luminosity AGNs into the epoch of reionization, the relatively small field of view of deep surveys \citep[compared to all-sky surveys, such as SDSS;][]{Atek+2025_glimpse, Eisenstein+2023_JADES, Rieke+2023_JADES} restricts the comoving volume sampled. 
Consequently, the discovery rate of intrinsically rare, massive galaxies ($M_* > 10^{10}\, M_\odot$) embedded within the massive dark matter halos ($M_{\rm DM}\gtrsim 10^{12}\,M_\odot$) at $z>6$ remains low \citep{deGraaff+2025_QG}. 
This volume limitation results in a paucity of spectroscopically confirmed massive host galaxies harboring accreting massive BHs, leaving the properties of these systems largely unexplored. 

In this work, we present deep and comprehensive follow-up observations of a unique red quasar, \targ, from JWST NIRCam and MIRI imaging to NIRSpec G140M+G395M, MIRI LRS+MRS spectroscopy.  
Originally identified in the GOODS-North field \citep{Fujimoto+2022}, the source has been spectroscopically confirmed at $z=7.1899$ via HST grism spectroscopy and the detection of \cii\ emission with NOEMA. 
The broadband SED, constrained by NOEMA dust continuum measurements, indicates that the quasar resides within a vigorous starburst host. 
Notably, the robust mid-infrared detection in Spitzer IRAC and MIPS bands confirms that \targ\ satisfies the canonical photometric selection criteria for red quasars \citep[e.g., ][]{Donley+2012_rQSO}. 
Furthermore, the non-detection in the deep ($\sim2$~Ms) Chandra X-ray observations implies high column densities, consistent with a Compton-thick highly-accreting system. 
These properties suggest that \targ\ likely represents a pivotal transition phase from a dusty starburst to an optically luminous quasar \citep[e.g.,][]{Sanders+1988, Hopkins+2008}, and might evolve into the luminous quasar at $z\sim 6$. 

This paper is organized as follows. 
In Section~\ref{sec2}, we present the observations and data reduction procedures. 
Section~\ref{sec3} describes the methodology used for the model fitting and spectral analysis. 
In Section~\ref{sec4}, we present the derived physical properties of both the AGN and the host galaxy. 
Finally, we summarize our conclusions in Section~\ref{sec5}.
Throughout this paper, we assume a flat universe with $\Omega_{\rm m}=0.3, \Omega_\Lambda=0.7,$ and $H_0=70\,\rm km\,s^{-1}\,Mpc^{-1}$. 
We use magnitudes in the AB system \citep{Oke+1983}. 

\section{Data}
\label{sec2}

\subsection{NIRCam image}
\label{sec2.1: NIRCam}

The NIRCam imaging data are retrieved from two programs: GO-4762 (PIs: S.~Fujimoto \& G.~Brammer) and the FRESCO survey (GO-1895; \citealt{Oesch+2023_FRESCO}). 
The source was observed with NIRCam F090W, F115W, F182M, F210M, F356W, and F444W filters. 
The detailed data reduction procedures are consistent with the standard data reduction procedure \cite[e.g., ][]{Oesch+2023_FRESCO, Eisenstein+2023_JADES, Atek+2025_glimpse}. 
Here we briefly mention the process.

The reduction of NIRCam imaging data commenced with the processing of uncalibrated files using the JWST Science Calibration Pipeline (v12.0.9) and CRDS context file \texttt{jwst\_1321.pmap}. 
Beyond standard STScI procedures, we implemented custom steps to mitigate artifacts, including the removal of snowballs, wisps, cosmic rays, persistence, and diffraction spikes, alongside an initial {1/\emph{f}} noise correction. 
Following Stage 2 pipeline processing, we applied a secondary correction for residual {1/\emph{f}} noise via sigma-clipping and performed two-dimensional background subtraction using the \texttt{sep} package. 
The data were then co-added through Stage 3 to produce final astrometrically aligned mosaics (referenced to Gaia stars) with pixel scales of 20 mas~pix$^{-1}$ for SW bands and 40 mas~pix$^{-1}$ for LW bands.
The NIRCam grism spectrum also confirms the presence of the broad component \citep{Oesch+2023_FRESCO}, we prioritize the NIRSpec data for our quantitative analysis due to its higher SNR.

\subsection{MIRI image}
\label{sec2.2}

The MIRI imaging data are also retrieved from two programs: F1280W from GO-4762 (PIs: S.~Fujimoto \& G.~Brammer) and F1000W and F2100W from the MEOW survey (GO-5407; \citealt{leung+2024}). 
The F1280W data were reduced following the standard procedures established by the Dawn JWST Archive (DJA; \citealt{Valentino+2023_QGs}) and widely used in public surveys \citep[e.g., PRIMER][]{Dunlop+2021_PRIMER}. Specifically, we retrieved the Level-2 products and processed them using the \texttt{grizili} software package \citep{Brammer+2021_grizili, Brammer+2022_grizili}, adopting the methodology described in detail by \cite{Valentino+2023_QGs}.

For F1000W and F2100W, the data publicly available in September 2025 for the GOODS-N field, jointly with the proprietary MEOW data, were reduced with the JWST Rainbow pipeline following the methodology described in \cite{Perez-Gonzalez+2024a_MIRI}. The observations come from PIDs 1181, 1264, 2926, 4762, and 5407, and mosaics were constructed for filters F560W, F770W, F1000W, F1280W, F1800W, and F2100W. The position of GNz7q is only covered at 10, 12.8, and 21~$\mu$m. Details about this reduction (jointly with a similar effort in other fields such as GOODS-S, EGS, or COSMOS) will be presented in P\'erez-Gonz\'alez et al. (in prep.). We briefly describe here the methodology. The JWST Rainbow database uses the official JWST pipeline adding some extra bespoke reduction step to deal with the background homogenization of the MIRI images. For this paper, the reduction was carried out with JWST version v1.19.41, reference files in jwst\_1413.pmap. 
A key component of this workflow is the `superbackground' strategy, in which background maps for individual exposures are constructed using the full set of campaign images, with known sources masked to mitigate bias. 
This approach yields a highly homogeneous background in terms of both level and noise. 
Quantitatively, this method reduces the standard deviation of background pixels in individual cal.fits files by a factor of $\sim$3.5 relative to the standard pipeline, resulting in a sensitivity improvement of 0.8 mag in the final mosaic. See Appendix A in \citealt{Perez-Gonzalez+2024a_MIRI} for further details, see also \citealt{2025A&A...696A..57O} and \citealt{2024ApJ...976..224A} reporting similar improvements in the noise of the final mosaics.

\subsection{NIRSpec data}
\label{sec2.3}

The deep spectroscopic follow-up of \targ\ was obtained with JWST/NIRSpec as part of the program (PID~4762l PIs: S.~Fujimoto \& G.~Brammer), using the MOS mode with the G395M grating in F290LP filter ($R\sim 1000$) and G140M grating in F140LP filter. 
We design the NIRSpec MSA configurations following a strategy similar to that adopted in the UNCOVER survey \citep{Bezanson+24, Price+25}, aiming both to secure the full integration on \targ\ and to maximize the multiplexing return of the deep G395M observations. 
To this end, we implement a two–tiered prioritization scheme distinguishing between (1) extraordinary or otherwise scientifically unique individual sources, and (2) samples of galaxies of broader interest (e.g., statistical sets of $z>4$ galaxies). 
Within this framework, the ranking order is set such that high–priority individual sources are placed first, followed by interesting but less extraordinary individual galaxies, and finally sample-based sources with equal priority within a given catalog. 
This scheme is used in combination with our custom MSA optimization algorithm to maximize slit allocation efficiency. 
As a result, a total of 241 sources are allocated slits. 
The on-source integration time for those sources is 7525~s ($\simeq2.1$~hr). 
The data were reduced with the \texttt{msaexp} pipeline \citep{Brammer2023} following standard procedures \citep[e.g.,][]{Heintz+25, deGraaff+25, Naidu+2025_BH*, Valentino+25}.

\subsection{MIRI MRS and LRS}

\targ\ was observed with the Medium Resolution Spectrograph (MRS; \citealt{Wells+15, Argyriou+23}) of the Mid-Infrared Instrument (MIRI; \citealt{Rieke+15, Wright+15, Wright+23}) on \targ\ as part of the cycle 3 JWST program GO-4762 (PIs: S.~Fujimoto \& G.~Brammer). The observations were performed with the MRS medium band that cover the wavelength ranges of 5.66$-$6.63\,$\mu$m, 8.67$-$10.13\,$\mu$m, 13.34$-$15.57\,$\mu$m, and  20.69$-$24.48\,$\mu$m for channels 1, 2, 3, and 4, respectively. We focus on channel 3 that covers the Pa-$\alpha$ emission line at a redshift of 7.2. The total on-source integration time corresponds to 11939 seconds (3.3\,hours), distributed in 8 dither positions that combine the positive and negative dither patterns optimized for point-like sources. For each dither position, a total of two integrations with 17 groups each were obtained using the SLOWR1 readout mode. 

The MRS observations were processed with version 1.17.1 of the JWST calibration pipeline and context 1322 of the Calibration Reference Data System (CRDS). We followed the standard MRS pipeline procedure,  
with additional customized steps to improve the quality of the final MRS calibrated products (see \citealt{Alvarez-Marquez+2023_dsfg, Alvarez-Marquez_2024} for details). The final channel 3 cube has a spatial and spectral sampling of 0.2"\,$\times$\,0.2"\,$\times$\,2.5\,nm \citep{Law2023}, respectively, and a resolving power of about 2500 \citep{Labiano2021, Jones2023}.

We extracted the 1D integrated spectra of \targ\ using a circular aperture with a radius equal to 0.5 arcsec for channel 3 (see Figure~\ref{fig4:miri}), respectively. We also extracted 15 1D background spectra using the same apertures at random positions of the MRS FoV clean of any emission of the source. We combined these spectra to generate the 1D median and standard deviation of the local background for channel 3. The median, which is compatible with zero, was subtracted from the \targ\ spectra with the goal of removing any systematic residual feature left in the MRS calibration process. The standard deviation was assumed to be the 1\,$\sigma$ uncertainty of \targ\ spectra. As \targ\ is considered spatially unresolved on the MRS observations, we implemented an aperture correction to derive the total Pa$\alpha$ spectra of \targ. For the selected apertures in channels 3, the aperture-correction factor is 1.59 following the MRS PSF (\citealt{Argyriou+23}). 

Data reduction and calibration for the MIRI LRS observations are conducted utilizing the standard JWST pipeline in conjunction with the \texttt{msaexp} \citep{Brammer2023}. 
One-dimensional spectral extraction was performed via the optimal extraction method \citep{Horne1986_spec}, employing a point-source profile derived from calibration observations of the A1V standard star BD+601753 \citep[Program ID: CAL-6620; ][]{Sloan_MIRI_cal_psf}. 
The 2D spectrum and the extracted 1D spectrum are presented in Figure~\ref{fig1.1: LRS}, and the data analysis is presented in Section~\ref{sec3.2: MIRI spec}. 

\section{Analysis}
\label{sec3}

\begin{deluxetable}{ccc}
\tablenum{1}
\tablecaption{Properties of \targ\ \label{tab1:source}}
    \tablewidth{0pt}
    \tablehead{
    \colhead{Properties} & \colhead{Value} & \colhead{Unit}
    }
    \decimalcolnumbers
    \startdata
    R.A. & 12:36:17.0\\
    Dec. & +62:12:32.27\\
    $z$ & 7.1899\\
    \hline
    & \textbf{Spectral fitting} \\
    $\log L_{\rm 5100\AA}$ & 45.02$\pm$0.01 & $\rm erg\,s^{-1}$\\
    $\alpha_1$ & 2.12$\pm$0.01 & \\
    $\alpha_2$ & 1.83$\pm$0.01 & \\
    $\rm FWHM_{\rm Fe}$ & 1039$\pm$14 & $\rm km\,s^{-1}$\\
    $f_{\rm Fe, 4434-4684\AA}$ & 1968$\pm$34 & $10^{-20}\,\rm erg\,s^{-1}\,cm^{-2}$ \\ 
    $\rm FWHM_{H\beta,BLR}$ & 2221$\pm$20 & $\rm km\,s^{-1}$\\
    $\rm FWHM_{H\beta,rBLR}$ & 11184$\pm$30 & $\rm km\,s^{-1}$\\
    $f_{\rm H\beta,BLR}$ & 932$\pm$30 & $10^{-20}\,\rm erg\,s^{-1}\,cm^{-2}$ \\
    $f_{\rm H\beta,rBLR}$ & 1381$\pm$100 & $10^{-20}\,\rm erg\,s^{-1}\,cm^{-2}$ \\
    $f_{\rm H\alpha,BLR}$ & 2080$\pm$87 & $10^{-20}\rm erg\,s^{-1}\,cm^{-2}$\\
    $f_{\rm H\alpha,rBLR}$ & 10843$\pm$172 & $10^{-20}\rm erg\,s^{-1}\,cm^{-2}$\\
    $f_{\rm H\gamma,BLR}$ & 366$\pm$21 & $10^{-20}\rm erg\,s^{-1}\,cm^{-2}$\\
    $f_{\rm H\gamma,rBLR}$ & 838$\pm$47 & $10^{-20}\rm erg\,s^{-1}\,cm^{-2}$\\
    $\rm FWHM_{[OIII]}$ & 193$\pm$8 & $\rm km\,s^{-1}$\\
    $f_{\rm [OIII]}$ & 138$\pm$9 & $10^{-20}\rm erg\,s^{-1}\,cm^{-2}$\\
    $f_{\rm [OIII],of}$ & 94$\pm$8 & $10^{-20}\rm erg\,s^{-1}\,cm^{-2}$\\
    $f_{\rm H\alpha,NLR}$ & 461$\pm$26 & $10^{-20}\rm erg\,s^{-1}\,cm^{-2}$\\
    $f_{\rm H\alpha,of}$ & 311$\pm$25 & $10^{-20}\rm erg\,s^{-1}\,cm^{-2}$\\
    $f_{\rm H\beta,NLR}$ & 157$\pm$12 & $10^{-20}\rm erg\,s^{-1}\,cm^{-2}$\\
    $f_{\rm H\beta,of}$ & 106$\pm$11 & $10^{-20}\rm erg\,s^{-1}\,cm^{-2}$\\
    $f_{\rm H\gamma,NLR}$ & 44$\pm$9 & $10^{-20}\rm erg\,s^{-1}\,cm^{-2}$\\
    $f_{\rm H\gamma,of}$ & 30$\pm$8 & $10^{-20}\rm erg\,s^{-1}\,cm^{-2}$\\
    \hline
    & \textbf{AGN properties} \\
    $\log M_{\rm BH}$ & 7.55$\pm$0.34 & $M_\odot$\\
    $\lambda_{\rm Edd}$ & 2.7$\pm$0.4 & \\
    $\log L_{\rm bol}$ & 46.2$\pm$0.1 & $\rm erg\,s^{-1}$\\
    $A_V$ & $0.21_{-0.08}^{+0.09}$ & mag \\
    \hline
    & \textbf{Host galaxy properties} \\
    $\log M_*$ & 10.5$\pm$0.4 & $M_\odot$ \\
    $\log L_{\rm FIR}$ & 12.3$\pm$0.3 & $L_\odot$ \\
    $\rm SFR_{Pa\alpha}$ & 250$\pm$88 & $M_\odot\,\rm yr^{-1}$ \\
    $\rm SFR_{FIR}$ & 330$\pm$97 & $M_\odot\,\rm yr^{-1}$ \\
    $\log L_{\rm [CII]}$ $^1$ & $(1.1\pm0.3)\times 10^9$ & $L_\odot$ \\
    $\rm FWHM_{\rm [CII]}$ $^1$ & $280\pm40$ & $\rm km\,s^{-1}$
    \enddata
    \tablecomments{Properties of this source, including the spectra and SED fitting result. \\
    $^1$ Values are adopted from \cite{Fujimoto+2022}. }
\end{deluxetable}

\begin{deluxetable}{ccc}
\tablenum{2}
\tablecaption{\targ\ properties from MIRI \label{tab1: MIRI}}
    \tablewidth{0pt}
    \tablehead{
    \colhead{Properties} & \colhead{Value} & \colhead{Unit}
    }
    \decimalcolnumbers
    \startdata
    & \textbf{Results from MIRI data} \\
    $f_{\rm HeI\lambda7281}$ & 1683$\pm$271 & $10^{-20}\rm erg\,s^{-1}\,cm^{-2}$\\
    $f_{\rm HeI\lambda 7816}$ & 2068$\pm$370 & $10^{-20}\rm erg\,s^{-1}\,cm^{-2}$\\
    $f_{\rm HeI\lambda 10031}$ & $<$280 & $10^{-20}\rm erg\,s^{-1}\,cm^{-2}$\\
    $f_{\rm HeI\lambda 10830}$ & 508$\pm$208 & $10^{-20}\rm erg\,s^{-1}\,cm^{-2}$\\
    $f_{\rm OI \lambda 8446}$ & 2179$\pm$303 & $10^{-20}\rm erg\,s^{-1}\,cm^{-2}$\\
    $f_{\rm Pa\gamma}$ & 427$\pm$212 & $10^{-20}\rm erg\,s^{-1}\,cm^{-2}$\\
    $f_{\rm Pa\beta}$ & 631$\pm$269 & $10^{-20}\rm erg\,s^{-1}\,cm^{-2}$\\
    $f_{\rm Pa\alpha}$ & 880$\pm$220 & $10^{-20}\rm erg\,s^{-1}\,cm^{-2}$
    \enddata
    \tablecomments{Derived physical properties of rest-frame near-infrared emission lines from the MIRI spectra. 
    Due to the instrumental broadening and limited spectral resolution, intrinsic line widths are omitted. 
    All non-detections are reported as $2\sigma$ upper limits.
    Given the poor spectral resolution of the MIRI LRS, no FWHM of those emission lines is shown. 
    }
\end{deluxetable}

\subsection{NIRSpec G395M spectrum \& fitting}
\label{sec3.1}

\begin{figure*}
    \centering
    \includegraphics[width=\linewidth]{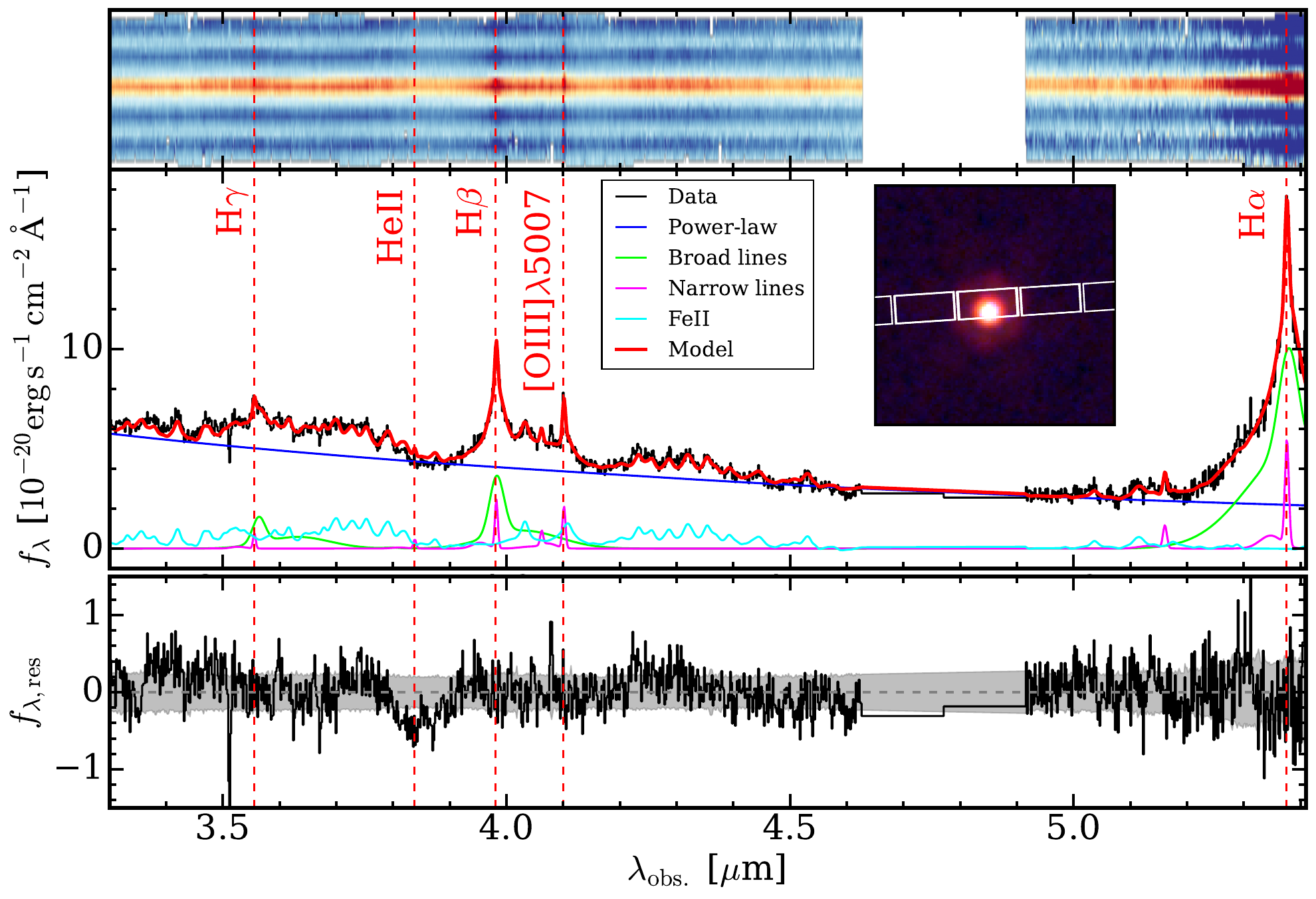}
    \caption{
    The NIRSpec/MSA spectrum and the best-fit model. 
    Top panel: The 2D spectrum generated from the JWST/NIRSpec observation. 
    Middle panel: The 1D spectrum (black line) extracted from the 2D spectrum and the best-fitting result (red line). The best-fit model consists of several components, including a featureless power-law continuum (blue line), an iron pseudo-continuum (cyan line), the narrow component of emission lines (magenta line), and the broad component of Balmer lines (green line). 
    Bottom panel: The residual between the data and the best-fit model (black line), and the uncertainties (gray dashed region).
    }
    \label{fig1:spec}
\end{figure*}

\begin{figure*}
    \centering
    \includegraphics[width=\linewidth]{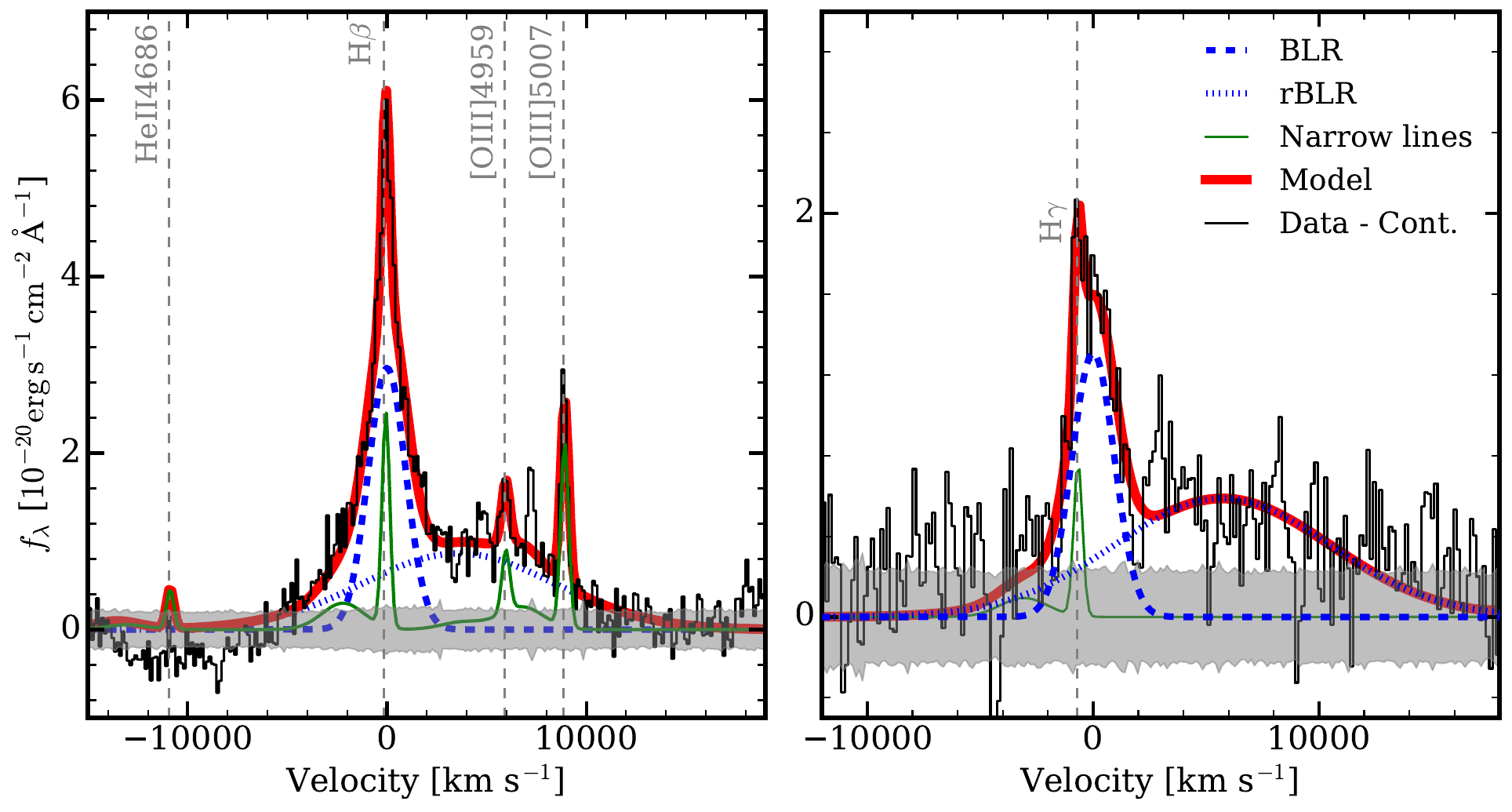}
    \caption{
    Zoom-in view of the \hb\ and \hg\ emission lines. 
    The black line shows the observational data after subtracting the global continuum (power-law and Iron Pseudo-continuum). 
    The best-fit model for the emission line is shown in red. 
    Individual components are color-coded, narrow emission is shown as the green line and the emission from the broad line region is shown as blue dashed lines. 
    The two components for the broad emission line are denoted as the blue-dashed and blue-dotted (rBLR) lines, respectively. 
    This redshifted component has been observed in some local quasars, in particular, quasars with high Eddington ratio \citep{Sulentic+2002_VBC, Marziani+2009}.
    }
    \label{fig2: zoomin}
\end{figure*}

We perform the spectral decomposition using \galfits\citep{Chen+2025_lrd, Chen+2025_lrd_gas, Li+2025_galfits}, a software developed for simultaneous SED, morphological, and spectral analysis, which can also be used to fit the spectra. 
The spectral fitting methodology implemented in \galfits\ is similar to that of established AGN analysis software \citep[e.g., \texttt{pyQSOFIT}, \texttt{fantasy};][]{Guo+2018_pyQSOfit, Ren+2024, Ilic+2020_fantasy}, principally decomposing the data into host galaxy continuum, AGN continuum, and emission line features. 
However, given the high luminosity of this source \citep{Yang+2023_qso}, the observed spectrum is dominated by the central AGN. 
We therefore consider the stellar continuum contribution negligible in the rest-frame optical and exclude it from the fitting process. 
This approximation is subsequently confirmed by the broadband SED analysis presented in Section~\ref{sec4.2: host} (see also Figure~\ref{fig3: sed}). 
Although broadband SED fitting indicates a significant host galaxy contribution within the MIRI filters, its impact in the rest-frame optical remains sub-dominant and is comparable to the local noise floor. 
To verify this, we performed a spectral decomposition using \texttt{pyQSOFIT}, incorporating both AGN and host components. 
The results are consistent with our primary analysis, confirming that the host galaxy contribution is negligible across the rest-frame optical regime.
Consequently, our final spectral model consists of two primary AGN components: a continuum model (comprising a double power-law and an Iron pseudo-continuum), and the emission line models.

The power-law continuum is parameterized by its amplitude ($f_0$), spectral index ($\alpha$), and a pivot wavelength ($\lambda_0$) fixed at 5000\,${\rm \AA}$ in the rest frame ($f_\lambda=f_0(\lambda/\lambda_0)^{-\alpha}$). 
For the iron pseudo-continuum, we employ the empirical \feii\ template derived from I\,Zw\,1 \citep{Boroson_Green1992}. 
The free parameters for this component are amplitude, velocity shift, and broadening velocity width.

The emission line model incorporates the permitted emission lines (\ha, \hb, \hg, and \heii) and the forbidden lines (\oiii\ $\lambda\lambda$4959, 5007 and \nii\ $\lambda\lambda$6548, 6584), assuming a tied line width across all species. 
We model the \oiii\ lines using two distinct components: a primary narrow-core component tracing the systemic velocity, and a secondary, broader, blueshifted component to account for potential gas outflows. 
During the fit, the flux ratio of the \oiii\ $\lambda$5007 to $\lambda$4959 doublet components is held fixed at its theoretical value of 2.98 \citep{Storey+2000_OIII}. The line ratio for the \nii\ $\lambda\lambda$6548,6584 doublets is also fixed to its theoretical value of 2.94.

The decomposed \oiii\ profile is then used to impose a set of physical constraints on the other emission lines. 
The profile of the \oiii\ lines is adopted as the template for all narrow-line components, i.e., the systemic velocity, the line width, and the line ratio between core and outflow component, are fixed when fitting the other forbidden lines and the narrow component of the Balmer lines. 
We adopted the \oiii\ doublet as the narrow-line template, as it is a standard tracer and represents the strongest, least-blended forbidden lines in our spectrum. 
Although the \sii\ doublet is another common choice, it falls outside the wavelength coverage of our data and is therefore not used in our fitting. 
We characterize the broad component of the Balmer lines using two distinct Gaussian components. 
This double-Gaussian model is necessary to characterize the shape of the spectrum. 
We will explain it in the next section and discuss the potential origin of the line profile in Section~\ref{sec4.1: Edd}. 
To mitigate the degeneracy between the broad Balmer line profiles and the Iron pseudo-continuum, the kinematic parameters (i.e., line width and velocity center) of each broad component are tied to be the same across the \ha, \hb, and \hg\ lines.

Using this model, we perform the fitting using \galfits\ with the \texttt{multi} task. 
This routine executes the fitting algorithm multiple times independently. 
For each iteration, a new set of initial parameters is randomly sampled from the pre-defined, physically allowed parameter space. 
The solution from this ensemble of fits that yields the global minimum reduced chi-squared ($\chi^2_{\nu}$) is then adopted as our final best-fit.

The results of our spectral decomposition are presented in Figure~\ref{fig1:spec}. 
The composite model provides an excellent representation of the observed rest-optical spectrum. 
All significant spectral features, including the various emission lines and the underlying Iron pseudo-continuum, are well-reproduced. 
The quality of the fit is further demonstrated by the residual spectrum (bottom panel in Figure~\ref{fig1:spec}), which shows no significant structured features and remains almost entirely within the 1$\sigma$ flux uncertainties. 

The best-fit result reveals that the broad Balmer lines can be well-characterized by two components, including a narrower component from the broad line region (BLR), whose velocity center is consistent with the systemic velocity of the narrow-line region, and a secondary redshifted component (redshifted-BLR, dubbed as rBLR; $\Delta v\approx 3600\,\rm km\,s^{-1}$) that has a much broader line width ($\rm FWHM\approx 11000\,km\,s^{-1}$), which have been observed in quasars at lower redshifts \citep[e.g., ][]{Sulentic+2002_VBC, Marziani+2009}. 
The origin of this redshifted, extremely broad component is still under debate, including the possibility of inflowing gas feeding the central BH growth, or a tidal disruption event \citep{Li+2023_tde}. 
We will discuss the reason for this redshifted broad component in a separate paper (Fujimoto et al. in prep.).
This two-component structure is required to fit both the \hb\ and \hg\ lines (Figure~\ref{fig2: zoomin}), and we further evaluate the necessity of this component in Section~\ref{sec3.2: eva}. 
While the same model is applied to \ha, the constraints on its redshifted component are less clear, as the far red wing extends beyond the spectrum's wavelength coverage.



\subsection{MIRI LRS \& MRS}
\label{sec3.2: MIRI spec}

\begin{figure*}
    \centering
    \includegraphics[width=0.46\linewidth]{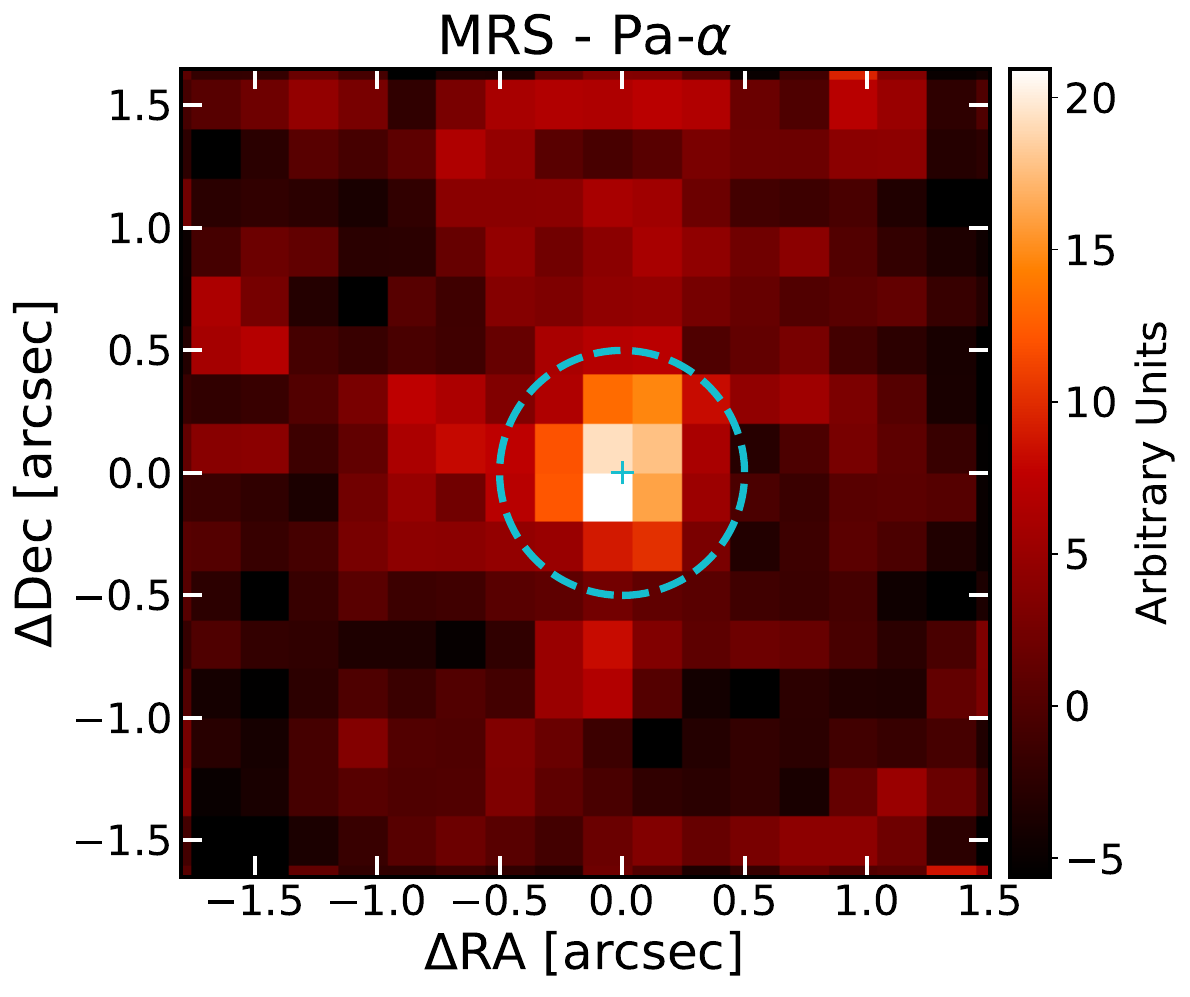}
    \includegraphics[width=0.53\linewidth]{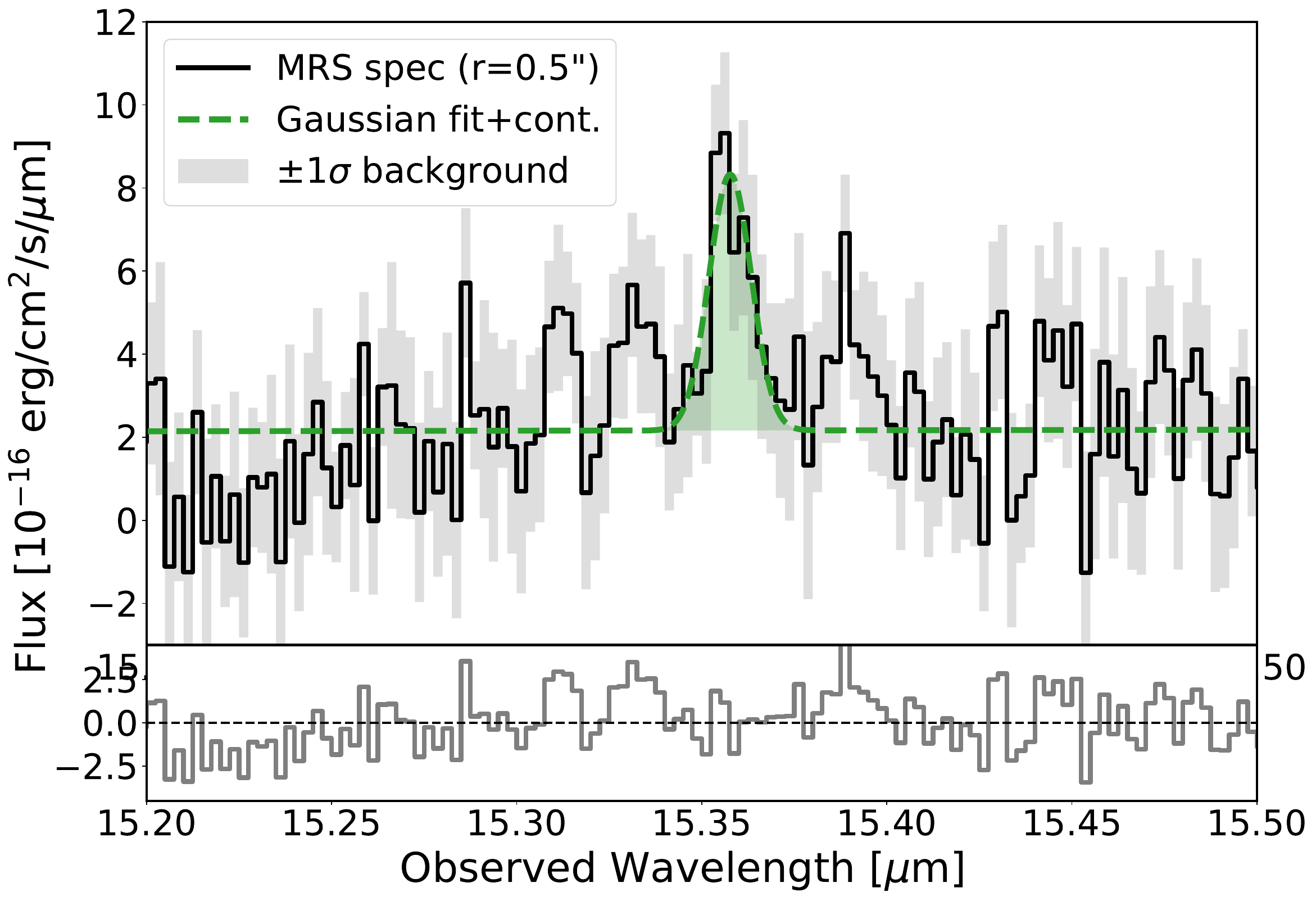}
    \caption{
    {\it Left:} The line intensity map generated from the MIRI MRS data around the Pa$\alpha$ emission, with a bandwidth of 1000\,$\rm km\,s^{-1}$. 
    The lightblue cross in the panel presents the locus of the quasar, and the dashed circle represents an aperture with a radius of 0\farcs5.
    {\it Right:} Spectrum extracted from the central $0\farcs5$ aperture of the MIRI observation. 
    Upper panel: The observed flux (solid black line) and associated error (grey shaded region), compared against the best-fit continuum and emission model (green line). 
    Lower panel: The residuals derived from subtracting the model from the observed data.}
    \label{fig4:miri}
\end{figure*}

The MIRI LRS 1D spectrum was modeled using a methodology consistent with our analysis of the NIRSpec G395M data, employing a model comprised of a continuum and Gaussian emission lines. 
A diverse suite of emission features was identified, including \textsc{H}e\textsc{i}, \textsc{Oi}, and the Paschen series. 
To isolate the continuum, we performed a preliminary fit using a single power-law after masking these emission regions.
As the stellar and dust torus components are not negligible in the observed wavelength regime (see Section~\ref{sec4.2: host}), we model the underlying continuum using a dual power-law formulation. 
In this framework, the first component characterizes the primary emission from the central engine, while the second accounts for the combined contributions from the host galaxy's stellar population and the thermal emission from the dust torus.

Subsequently, we model the identified emission features using the continuum-subtracted spectrum. Due to the limited spectral resolution of the MIRI LRS instrument \citep{Kendrew+2015_MIRI_LRS_spec_res}, a detailed decomposition into narrow and broad components is generally omitted in favor of single Gaussian profiles to characterize line morphologies. 
An exception was made for the \ha\ complex, which is modeled with a double Gaussian profile utilizing empirical constraints derived from our NIRSpec observations.
Parameter optimization was conducted via the \texttt{lmfit} Python package \citep{Newville+2014_lmfit}. 
To maintain maximal flexibility amidst the instrument's resolution limits, we allow the centroid, width, and amplitude of each line to vary independently rather than imposing tied kinematic constraints. 
This multi-stage approach culminated in a global fit of the MIRI LRS spectrum, employing a composite model of a continuum and Gaussian emission lines whose parameters are initialized from the results of the isolated fits. 
The final spectral decomposition is illustrated in Figure~\ref{fig1.1: LRS}, with the derived physical parameters shown in Table~\ref{tab1: MIRI}.

Our spectral decomposition reveals that emission features at shorter rest-frame wavelengths, specifically \textsc{H}e\textsc{I}$\lambda$7281, \textsc{H}e\textsc{I}$\lambda$7816, and \textsc{OI}$\lambda$8446, exhibit marginally broader profiles relative to the Paschen series. 
However, we cannot definitively exclude the influence of instrumental broadening on these observed widths. 
The MIRI LRS resolving power is lower at the blue end of the spectrum ($R \sim 40$ at 5$\mu$m) compared to the red end ($R \sim 160$ at 10$\mu$m). 
Consequently, the current spectral resolution precludes a detailed kinematic analysis of these rest-frame near-infrared line profiles. 

For the MIRI MRS data, we perform a one-component Gaussian function to characterize the Pa-$\alpha$ emission line profiles, with a linear function to fit the background (see Figure~\ref{fig4:miri}). 
The final uncertainties in the fit parameters, such as FWHM, flux, and redshift, were obtained as the standard deviation of all the individual measurements of 1000 bootstrapped spectra after adding a random Gaussian noise equal to the RMS to the original spectrum. 
The integrated Pa-$\alpha$ flux is (8.8\,$\pm$\,2.2)$\times$10$^{-18}\,\rm erg\,s^{-1}\,cm^{-2}$ with an intrinsic FWHM of 230\,$\pm$\,100\,km\,s$^{-1}$. 
The redshift of the Pa-$\alpha$ line is 7.1881\,$\pm$\,0.0009 compatible with the one obtained from NIRSpec spectroscopy (See Table~\ref{tab1: MIRI})


\subsection{SED \& Morphological fitting}
\label{sec3.2}

We perform a simultaneous AGN-host galaxy decomposition and SED fitting using \galfits\ to study the host galaxy properties for \targ. 
The software employs a forward modeling approach to simultaneously characterize heterogeneous imaging data, accommodating different instruments, filters, and sensitivities (Li et al. in prep.).  
The methodology is identical to the procedure established in recent work \citep{Li+2025_galfits, Chen+2025_lrd, Chen+2025_lrd_gas}, which already successfully decomposes the AGN and galaxy component for high-$z$ AGNs. 
As the detailed setup and fitting process are fully described in \citet{Li+2025_galfits}, we here provide a brief overview of the setup for reference.

We firstly construct the empirical PSFs for NIRCam and MIRI images using the field stars \citep{Zhuang+2023_AGN, Li+2025_galfits}, which gives the best description for individual stars compared to the relatively sharper PSF generated from the \texttt{WebbPSF} \citep{Yue+2024_QSO}. 
The empirical PSF is then used for characterizing the unresolved nuclear emission from the quasar. 
For its nuclear SED, we adopt a non-parametric approach, treating the luminosity in each photometric band as an independent, free parameter. 


This PSF is also used for the host galaxy characterization. 
The host galaxy component is defined by a simultaneous fit to its morphological and spectral energy distribution (SED) properties. 
The morphology is parameterized by a single \sersic\ profile, which is defined by its centroid, \sersic\ index ($n$), effective radius ($R_e$), axis ratio ($q$), and position angle ($\theta$). 
The host's stellar SED is modeled via population synthesis assuming a \citet{Kroupa2001} initial mass function. 
We adopt a non-parametric star formation history (SFH), a choice motivated by the complex, bursty SFHs recently revealed in high-$z$ galaxies by JWST \citep{Wang+2024_rubies, Wright+2024}. 
The free parameters of this stellar population model include the total stellar mass ($M_*$), the star formation rate (SFR) within discrete temporal bins, the stellar metallicity ($Z$), and the V-band dust extinction ($A_V$), for which we assume a \citet{Calzetti+2000} attenuation law. 
We made a simple assumption of energy balance that the FIR energy comes from the stellar light attenuation, and therefore, implementing the FIR constraints from the JCMT and NOEMA observations \citep{Fujimoto+2022}.

To ensure robust convergence and avoid unphysical solutions, particularly in datasets with limited photometric coverage, the fitting process incorporates several priors. 
These include constraints from established mass-size relations, mass-dust extinction, and mass-metallicity relations, in addition to an SFH prior \citep{Allen+2025, Maheson+2024, Nakajima+2023}. 

We perform the fitting process using the nested-sampling algorithm \citep{Skilling2004}. 
We initiated the decomposition by fitting the multi-band imaging with only a pure PSF component. This model proved insufficient, as the residual images clearly revealed a significant, extended component spatially offset from the nucleus. 
Motivated by this, we tested a two-component model comprising a PSF and a \sersic\ component, allowing their centers to be fit independently. 
This model, which converged on the $\sim$2 kpc offset, resulted in a substantially lower BIC, confirming the spatial misalignment between the quasar and this extended host. 
Finally, we tested a more complex, three-component model to determine if a host component also exists co-spatially with the nucleus. 
This final model includes a PSF, a \sersic\ component co-centered with the PSF, and an offset \sersic\ component. 
This three-component configuration yielded the global minimum BIC, indicating it is the statistically preferred description. 

While our primary morphological and spectral analysis utilized \galfits, we employed \cigale~\citep{Boquien+2019_cigale} as an independent consistency check to validate the derived parameters. 
The detailed SED fitting process and the setup will be shown in Section~\ref{sec4.2: host}. 

\begin{figure*}
    \centering
    \includegraphics[width=\linewidth]{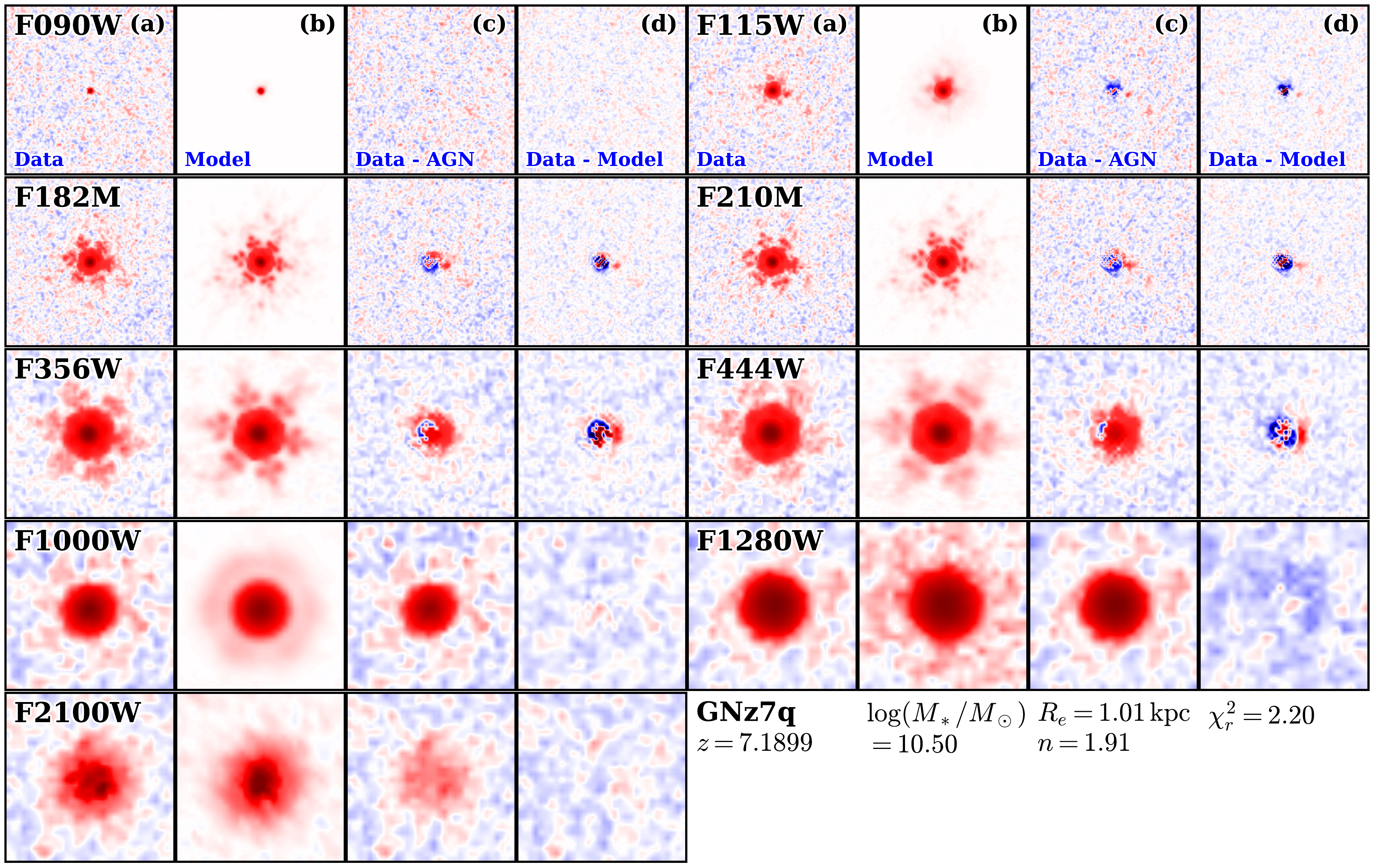}
    \caption{
    Multi-band AGN-host decomposition for \targ. 
    From left to right, the panels display: (a) the original JWST image; (b) the best-fit model; (c) the image of the host galaxy component after Point Spread Function (PSF) subtraction; and (d) the final residual map.
    }
    \label{fig2: galfits}
\end{figure*}

\subsection{Dust Extinction} 
\label{sec4.1: Av}

\cite{Fujimoto+2022} characterized \targ\ as a high-redshift red quasar hosted by a dusty starburst, indicating that the observed emission from both the nuclear and stellar components is strongly suppressed by dust. 
With the new NIRSpec observations, we therefore quantify the dust attenuation ($A_V$). This derived extinction is subsequently applied to infer the intrinsic black hole mass and the bolometric luminosity.

In the same manner as other high-$z$ AGNs studies \citep[e.g., ][]{Harikane+2023}, we estimate the dust attenuation for the quasar component via the Balmer decrement of the narrow emission lines (including the core component and the outflowing component; Table~\ref{tab1:source}). 
Adopting a typical dust extinction law and an intrinsic, dust-free narrow-line ratio of $f_{\rm H\alpha}/f_{\rm H\beta}=$2.86 \citep{Calzetti+2000, Osterbrock1989_book}, the observed ratio of $3.03_{-0.07}^{+0.08}$ yields an extinction of $A_V=0.21_{-0.08}^{+0.09}$~mag, following the method proposed in \cite{Dominguez+2013_dust} and \cite{Momcheva+2013_dust}. 


To validate our extinction estimate, we performed an independent estimation based on continuum fitting. 
We modeled the observed rest-frame UV-to-optical SED using a composite quasar template \citep{Selsing+2016_qso} reddened by a standard dust extinction law \citep{Calzetti+2000}, treating the $A_V$ as a free parameter. 
The resulting $A_V$ is consistent with the value derived from the Balmer decrement of the narrow emission line, while it is different from that obtained from the Balmer decrement derived from the broad emission line ($A_V=1.9$, assuming an intrinsic ratio of 3.06, which is a common value for local AGN; \citealt{Dong+2008_BLR}). 
However, caution is warranted as the high gas densities characteristic of the BLR deviate from standard Case B recombination assumptions. 
Consequently, the elevated $A_V$ derived from broad lines likely caused by the collisional excitation rather than the dust attenuation \citep{Son+2025_BLR, Yan+2025_Av}.

While Paschen-to-Balmer line ratios can offer an independent estimation for dust attenuation \citep{Reddy+2023_Paline}, the application in this analysis is severely limited by the spectral resolution. 
Specifically, the low spectral resolution of MIRI LRS precludes the decomposition of Pa$\beta$ and Pa$\gamma$ into distinct broad (AGN) and narrow (host) kinematic components, a challenge further compounded by the blending of Pa$\gamma$ with He\textsc{i},$\lambda$10830. 
In contrast, the MIRI MRS data recover only the narrow Pa$\alpha$ component attributed to the host galaxy. 
Consequently, deriving a robust $A_V$ from these inconsistent spectral components is non-trivial. A
lthough the observed Pa$\alpha$/Pa$\beta$ flux ratio is consistent with measurements of the broad line region in high-$z$ quasars \citep{Bosman+2024_qsoBLR, Bosman+2025_qso}, we caution that this agreement may be fortuitous given the mixing of physical origins. 
Definitive constraints on the extinction will require deeper, high-resolution spectroscopy capable of resolving the underlying kinematic substructure. 

The $A_V$ of the AGN can also be constrained by photometrically comparing the rest-frame UV--optical SED shape of \targ\ with standard quasar templates \citep{Kato+2020_rQSO, Selsing+2016_qso, Yang+2021}. 
We also employed this approach to estimate the intrinsic dust reddening. 
We note that, while the shape of the rest-frame optical continuum of \targ\ is similar to that of typical high-$z$ quasars, the rest-frame UV continuum exhibits a divergent slope distinct from the standard population. 
This spectral discrepancy suggests a non-standard extinction curve for the central AGN, which cannot be well modeled with a typical extinction curve, \citep[e.g., ][]{Calzetti+2000, Calzetti2001_ext, Gordon+2003_ext}. 
Such properties have been observed in other high-$z$ systems \citep{Gallerani+2010_dust, Sanders+2025_dust}, which may be attributable to specific dust formation mechanisms in quasars \citep{Maiolino+2004_dust, DiMascia+2021_dust}. 
A comprehensive analysis of this anomalous extinction curve is beyond the scope of the current work and will be presented in a separate paper (Fei et al. in prep.).
Consequently, we adopt the value derived from the narrow emission lines, $A_V=0.21_{-0.08}^{+0.09}$, as the fiducial extinction parameter for the subsequent analyses. 
This is the most conservative value in terms of the bolometric luminosity and BH mass estimate. 

\section{Results \& Discussion}
\label{sec4}

\subsection{BH properties}
\label{sec4.1: Edd}

We then estimate the black hole (BH) properties using the parameters derived from the best-fit model for the spectrum. 
In particular, the $M_{\rm BH}$ of an AGN is calculated using the virial estimator:
\begin{align}
M_{\rm BH}=f_{\rm BLR}\frac{R_{\rm BLR}\Delta V^2}{G},
\end{align}
where $G$ is the gravitational constant, $R_{\rm BLR}$ is the radius of the broad line region (BLR), and $f_{\rm BLR}$ is the virial factor, a dimensionless coefficient that accounts for the unknown geometry, kinematics, and inclination angle of the BLR. 
The velocity, $\Delta V$, is proxied using the width of the broad emission line (e.g., FWHM).
The measurement of $R_{\rm BLR}$ is conducted by reverberation mapping (RM), which measures the time lag between variations in the continuum flux and the subsequent response of the broad emission lines. 
We adopt a virial factor of $f_{\rm BLR}=4.16$, based on the reverberation mapping of SDSS quasars \citep{Shen+2024_RM}, assuming that the BLR geometry of \targ\ is consistent with that of its low-redshift analogs.
These RM studies have demonstrated a robust correlation between the BLR size ($R_{\rm BLR}$) and the AGN's luminosity. 
The luminosity is typically quantified by either the 5100\,${\rm \AA}$ monochromatic luminosity ($L_{\rm 5100}$) or the broad line luminosity ($L_{\rm BLR}$). 
This established Radius-Luminosity relationship is foundational, and it allows for the calibration of single-epoch virial mass estimators. 
These empirical formulae, which are benchmarked against AGN samples in the local Universe \citep[e.g., ][]{Greene+2005, Vestergaard+2006, Reines+2015}, enable the estimation of $M_{\rm BH}$ using the line widths and luminosities measured from a single spectrum even for high-$z$ AGNs. 
However, we caution that recent interferometric studies of luminous high-$z$ quasars suggest that the BLRs of super-Eddington accretors may be systematically larger than those of local quasars, potentially leading to an overestimation of $M_{\rm BH}$ by $\sim$0.3 dex \citep[e.g.,][]{Abuter+2024_BLR_dyn}. 

Initially identified as a Compton-thick, super-Eddington candidate exhibiting significant X-ray faintness \citep{Fujimoto+2022}, this source suggests that BH mass estimators derived from local, lower-accretion AGNs may require modification. 
The reverberation mapping analyses have revealed that the canonical luminosity-size relation, which is calibrated on local AGNs with moderate accretion rates, is not universally applicable \citep{Du+2015, Du+2016, Du+2018} and is regulated by the accretion rate. 
This is caused by the theoretical prediction that the canonical BLR framework from the photo-ionizing model, $R_{\rm BLR}\propto L_{\rm ion}^{1/2}$, where $R_{\rm BLR}$ is the size of the BLR and $L_{\rm ion}$ is the luminosity of the ionizing source, requires specific structure of the BLR, which may be modified when the accretion rate exceeds the Eddington ratio \citep{Wang+2014c, Czerny+2019}. 
Studies targeting massive BHs with higher accretion rates in the local Universe have found that these objects exhibit \hb\ time lags that are systematically shorter by factors of $\sim$2–8 than predicted by the canonical relation for given luminosities \citep{Du+2015, Du+2016, Du+2018}. 
The statistical analysis of these local AGNs confirmed that the magnitude of this lag shortening strongly correlates with the accretion rate \citep{Du+2016, Du+2018}. 
A physical explanation for this discrepancy is the self-shadowing effect of a slim accretion disk, which is expected to form in super-Eddington regimes \citep[e.g., ][]{Wang+2014c}. 
In this model, the inner part of the disk becomes geometrically thick due to intense radiation pressure. 
This inflated disk then anisotropically shields the BLR from ionizing photons, causing the \hb\ ionization front to shrink. 
Given this established dependency, applying the canonical $R_{\text{BLR}}-L_{5100}$ relation to a high-accretion-rate object would result in an erroneous $R_{\text{BLR}}$ and, consequently, an incorrect BH mass \citep{Abuter+2024_BLR_dyn}. 

\begin{figure*}
    \centering
    \includegraphics[width=0.7\linewidth]{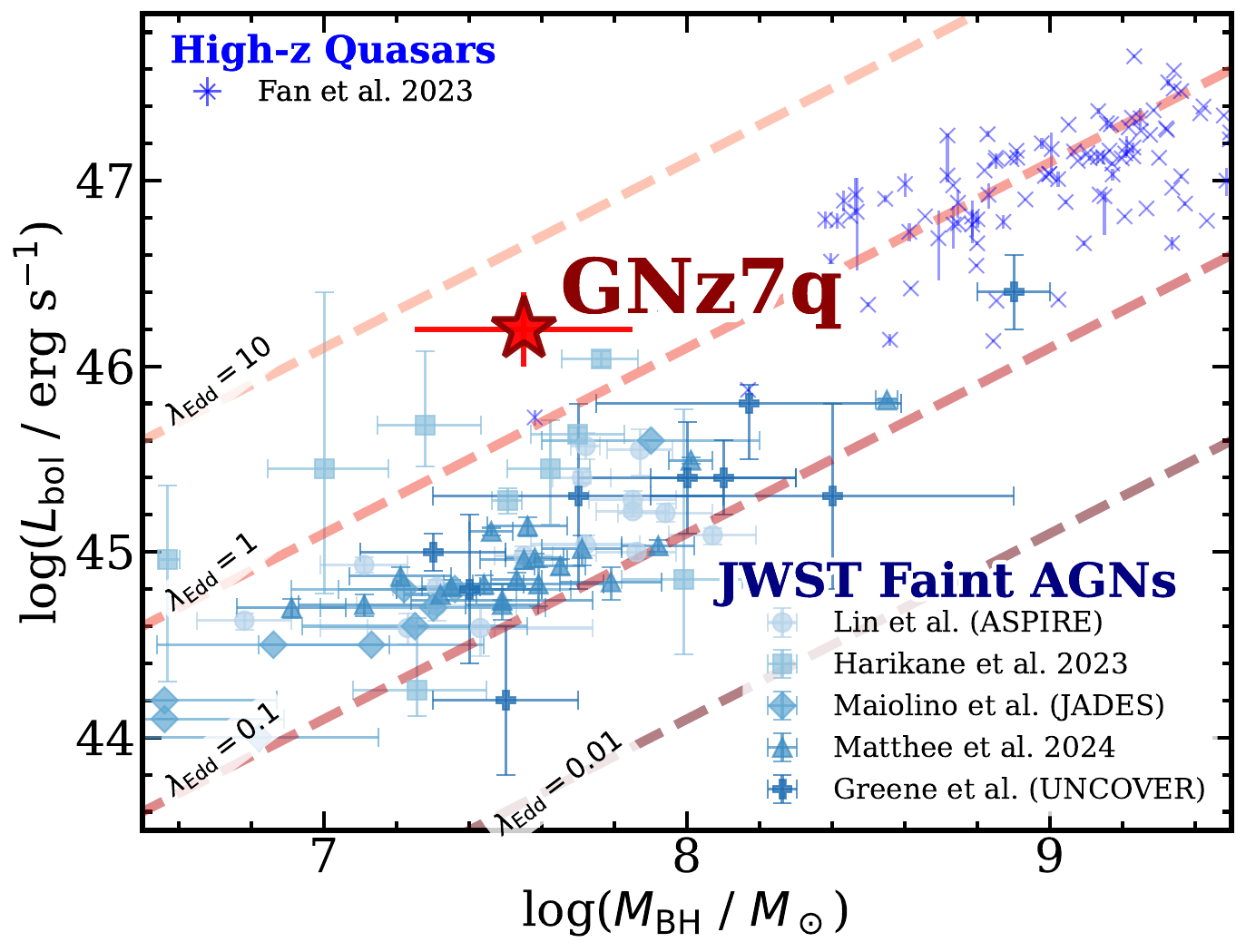}
    \caption{BH mass vs. bolometric luminosity of \targ\ and other AGNs at $z>4$, including the quasars discovered by ground-based facilities \citep{Shen+2019, Reed+2019, Yang+2021, Farina+2022} and the JWST discovered faint AGNs \citep{Matthee+2024_LRD, Harikane+2023, Maiolino+2024, Lin+2024_BHE, Greene+2024}.}
    \label{fig4: edd}
\end{figure*}

To estimate the BH mass, we first determine the BLR size using Equation (6) of \cite{Du+2015} with the \hb\ based empirical relationship, which explicitly accounts for the Eddington ratio. 
We then calculate the BH mass using this BLR size, a typical virial parameter, and an assumed high accretion rate of $\dot{\mathcal{M}}\geq 3$ \citep{Fujimoto+2022}. 
This procedure, incorporating the dust extinction correction of $A_V=0.21$, yields a mass for \targ\ of $\log(M_{\rm BH}/M_\odot) = 7.55\pm0.34$. The uncertainty is mainly contributed by the systematic uncertainty embedded within the calibrator, which is 0.31\,dex \citep{Shen+2024_RM}. 
This value is approximately two orders of magnitude lower than typical quasars at $z\gtrsim 6$ \citep{Fan+2023} and is consistent with the high-mass end of the BH mass function (BHMF) for the faint JWST-detected AGN population \citep{Matthee+2024_LRD, Taylor+2025_AGN}.

To validate our results, we cross-check the BH mass using two independent empirical relations. 
First, applying the continuum-based estimator (Equation 5 of \citealp{Du+2018}) yields $\log(M_{\rm BH}/M_\odot) = 7.60\pm0.40$. 
Second, using the \feii-based size-luminosity relation \citep{Du+2019, Pan+2025_DESI} results in $\log(M_{\rm BH}/M_\odot)=7.55\pm0.40$. 
Both values are in agreement with our primary Eddington-ratio corrected H$\beta$ estimate. 
This consistency indicates that the scaling between the BLR and continuum luminosities follows the empirical relationships established for the local Universe \citep{Greene+2005}, suggesting that \targ\ possesses a BLR structure comparable to typical low-redshift AGNs \citep{Bosman+2024_qsoBLR, Bosman+2025_qso}. 
We note that estimates using traditional scaling relations (excluding Eddington-ratio corrections, e.g., \citealt{Greene+2005}) also fall within the consistent range of $10^{7.5}$--$10^{7.8}\,M_\odot$, using different BH mass estimators. 
Given that the ionizing accretion disk is spatially compact relative to the BLR radius, we adopt the Eddington-ratio corrected, H$\beta$-based determination as our fiducial mass. 

The above-mentioned BH mass estimate already considers the effect of the dust extinction. 
Given the relatively low $A_V$ in the rest-frame optical ($A_V=0.21_{-0.09}^{+0.08}$), 
the corrected broad H$\beta$ flux of $f_{\rm H\beta,BLR} = 1039 \pm 30 \times 10^{-20}\,\rm erg\,s^{-1}\,cm^{-2}$. 
Due to the shallow dependence of the virial mass on luminosity ($M_{\rm BH} \propto L^{0.5}$), this correction corresponds to a marginal increase in the BH mass of approximately 5\% ($\sim 0.02$ dex), which falls well within the systematic uncertainties. 

We derive a bolometric luminosity of $L_{\rm bol}=(1.50 \pm 0.20)\times 10^{46}\,\rm erg\,s^{-1}$ from the best-fit SED. Taking uncertainties into account, this is in good agreement with the value of $(1.0\pm 0.3)\times 10^{46}\,\rm erg\,s^{-1}$ inferred from the 5100\,\AA\ monochromatic luminosity using the bolometric correction of \citet{Kaspi+2000}.
The Eddington ratio ($\lambda_{\rm Edd}$) was then calculated with \cite{Trakhtenbrot+2011}:
\begin{equation}
    \lambda_{\rm Edd}=\frac{L_{\rm bol}/\rm erg\,s^{-1}}{1.5\times 10^{38}M_{\rm BH}/M_\odot},
\end{equation}
where $1.5\times 10^{38}M_{\rm BH}/M_\odot$ is the Eddington luminosity of the BH. 
Based on the derived luminosity and the fiducial BH mass, the source is characterized by a highly super-Eddington ratio of $\lambda_{\rm Edd} =2.7\pm0.4$. 

Evidence for super-Eddington accretion is further strengthened by the identification of the \textsc{OI}$\lambda$8446 emission (Figure~\ref{fig1.1: LRS}; Table~\ref{tab1: MIRI}). 
As detailed by \cite{Inayoshi+2022_qso}, this low-ionization feature primarily originates from a dense gaseous disk ($0.1 \lesssim r \lesssim 1$pc) close to the central massive BH. 
Such broad \oi\ lines powered by AGNs were also discovered in recent faint AGNs (LRDs) discovered by JWST \citep{Tripodi+2025_BLR_LRD, Kokorev+2025_BH*, Sun+2026_LRD_BH*}

In Figure~\ref{fig4: edd}, we show the BH mass and bolometric luminosity relation of \targ\ by plotting its location on the $L_{\rm bol} - M_{\rm BH}$ plane alongside other AGNs in the early Universe ($z>4$). 
For comparison, we also show highly luminous quasars so far discovered with ground-based facilities and the emerging population of faint AGNs recently identified with JWST, at similar redshifts with \targ\ ($z>4$) compiled from the literature \citep{Shen+2019, Yang+2021, Farina+2022, Matthee+2024_LRD, Harikane+2023, Maiolino+2024, Lin+2024_BHE, Greene+2024}. 
Our analysis demonstrates that \targ\ is placed in the highest $\lambda_{\rm Edd}$ regime among the quasar/AGN populations so far identified at $z>4$. 
This significantly elevated accretion rate implies that the source is undergoing a phase of rapid mass assembly, consistent with the evolutionary pathway required to form the most massive SMBHs observed at $z \sim 6$.


\subsection{X-ray faintness}
\label{sec4.2: x-ray}
\begin{figure}
    \centering
    \includegraphics[width=\linewidth]{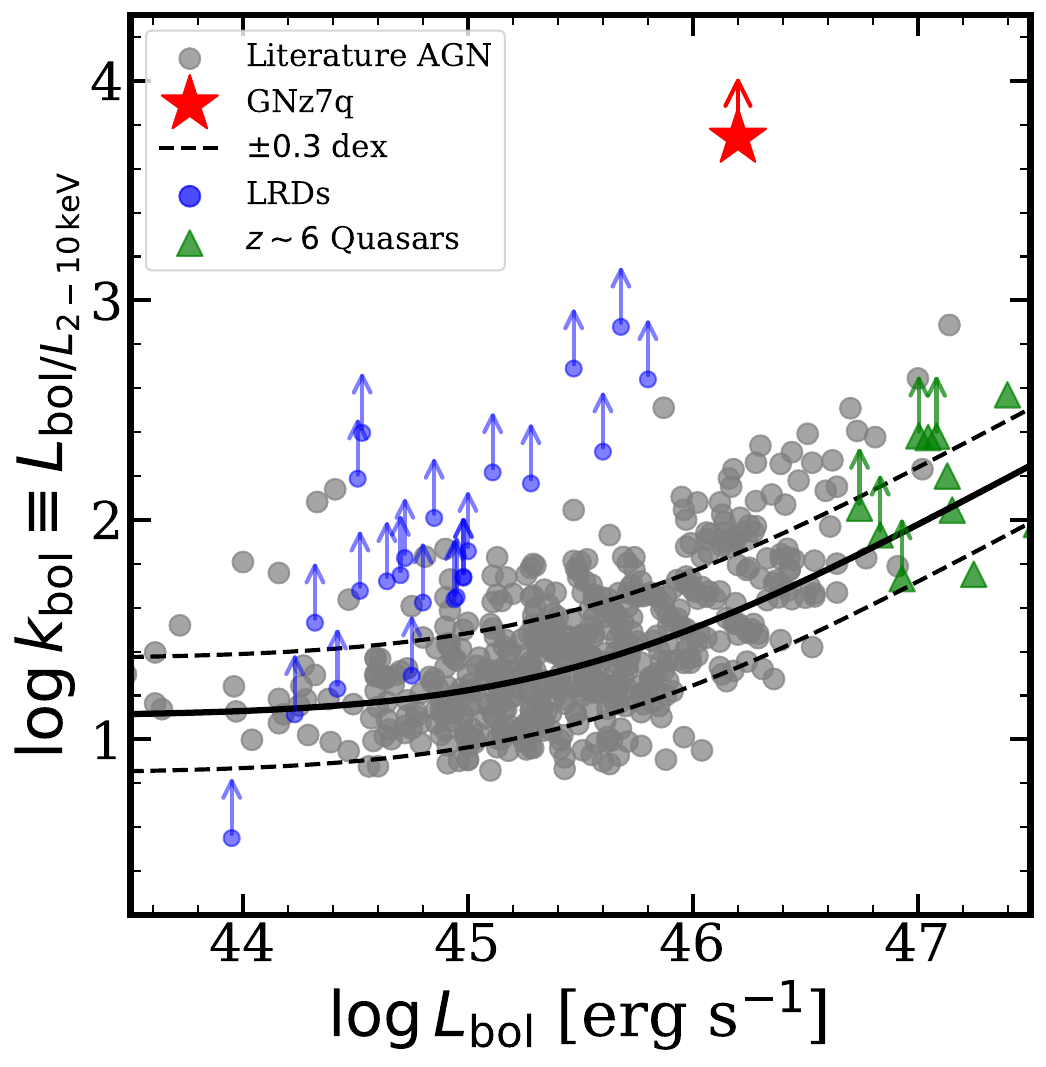}
    \caption{Unique X-ray faintness of \targ, compared to other AGNs in the literature. 
    \targ\ is shown as a red star. 
    Comparison samples include $z\sim 2$ quasars \citep[grey points;][]{Lusso+2010_xray}, high-$z$ Type 1 faint AGNs, mainly little red dots \citep[blue circles;][]{Maiolino+2025_xray}, and luminous quasars at $z\gtrsim 6$ \citep[green triangles;][]{Wang+2021_xray}. 
    The black solid line shows the empirical relationship between $k_{\rm bol}$ and $L_{\rm bol}$ from \cite{Duras+2020_xray}.}
    \label{fig4: x-ray}
\end{figure}

The nature of \targ\ as a candidate for a Compton-thick quasar at $z>7$ was initially suggested by its non-detection in one of the deepest Chandra X-ray surveys, despite its high luminosity in the infrared \citep{Fujimoto+2022}. 
This phenomenon of X-ray faintness has recently become a recurring theme in JWST observations, particularly among the emerging population of LRDs and faint broad-line AGNs \citep[e.g.,][]{Maiolino+2025_xray}. 
Motivated by this context, we revisit the X-ray properties of \targ\ utilizing our updated bolometric luminosity ($L_{\rm bol}$) estimate. 
Adopting an intrinsic X-ray photon index of $\Gamma=2.0$ that follows the relation $N(E)\propto E^{-\Gamma}$, where $E$ is the energy of photons and $N(E)$ represents the X-ray spectra, we derive an upper limit of $L_{\rm 2-10\,keV}<2.9\times 10^{42}\,\rm erg\,s^{-1}$. 
Figure~\ref{fig4: x-ray} compares \targ\ against the empirical relation between the bolometric to X-ray luminosity ratio ($k_{\rm bol}\equiv L_{\rm bol}/L_{2-10\,\rm keV}$) and bolometric luminosities \cite{Duras+2020_xray}. 
We also present other high-$z$ AGNs as the comparison sample \citep[e.g., ][]{Wang+2021_xray, Maiolino+2025_xray}. 
We find that \targ\ deviates from standard quasar scaling relations at a confidence level of $>5\sigma$. 
While this suppression generally mirrors the X-ray faint AGN population identified by JWST, \targ\ is far more extreme, exhibiting a deviation $\sim$1 dex deeper than even this faint population. 
This profound faintness reinforces the link between \targ\ and the most heavily obscured, rapidly growing BHs in the early Universe

Two physical mechanisms may account for this observed X-ray weakness. 
First, intrinsic X-ray production may be suppressed by the super-Eddington accretion flow. 
In this regime, the inner accretion disk becomes geometrically thick (`slim disk'), which can anisotropically shield the X-ray corona or prevent the formation of a standard reprocessing region due to self-shadowing effects \citep{Madau+2024_Xweak, Inayoshi+2025_Xweak_LRD}. 
Alternatively, the suppression may be extrinsic, resulting from heavy obscuration by a Compton-thick column ($N_{\rm H}\approx 10^{24}\,\rm cm^{-2}$) that blocks line-of-sight X-rays \citep{Maiolino+2025_xray}. 
Given the source's high Eddington ratio and its location within a dusty starburst host, it is plausible that a combination of both intrinsic super-Eddington suppression and high column density obscuration contributes to the observed properties.




\subsection{Host galaxy properties}
\label{sec4.2: host}

\begin{figure}
    \centering
    \includegraphics[width=\linewidth]{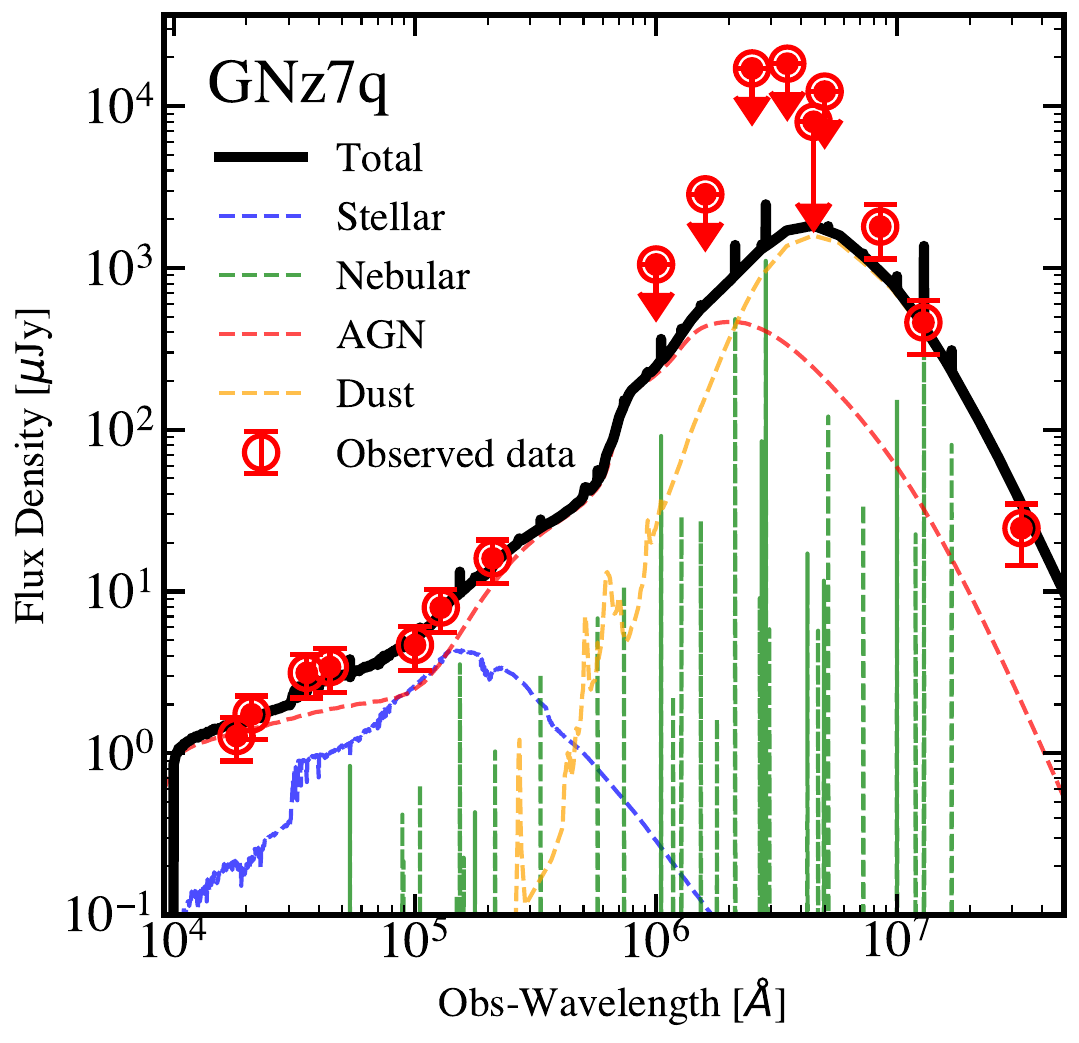}
    \caption{
    Best-fit multi-component SED decomposition obtained using \cigale. 
    The observed photometry is shown as red points, comprising both the new JWST observations and archival data \citep{Fujimoto+2022}. 
    Dashed lines indicate the individual model components, including stellar population (blue), nebular emission (green), AGN (red), and cold dust (yellow). 
    The solid black line represents the total integrated emission model.
    }
    \label{fig3: sed}
\end{figure}

Figure~\ref{fig2: galfits} presents the results of our multi-band photometric decomposition using \galfits. 
Crucially, the decomposition yields a clear and robust detection of the host galaxy, particularly in the F356W, F444W, F1000W, and F1280W filters where the stellar continuum is prominent. 
While the host is detected across the full instrument suite (with the exception of the F090W and F115W bands), these rest-frame optical-to-IR bands provide good constraints on the host galaxy emission, allowing the estimation of the host galaxy properties. 

Interestingly, the decomposition resolves two components in the rest-frame UV bands, a central component co-located with the quasar, and an off-center component situated approximately $2\,\text{kpc}$ away from the mass center determined from the long-wavelength filter. 
These clumps are similar to features seen in high-$z$ galaxies \citep[e.g., ][]{Tadaki+18_SMG, Gillman+2024_dsfg}, suggesting they may represent the star-forming regions in the early Universe. 
We also obtain a marginal detection in the F2100W filter (SNR$\approx$2). 
At this rest-frame wavelength ($\sim2.5\,\rm \mu m$), the emission is dominated by the hot dust from the AGN, consistent with the diminishing contribution from the stellar population. 


To accommodate both AGN and host galaxy components in \targ, we also use the \cigale\ for our SED fitting \citep{Boquien+2019_cigale, Yang+2022_cigale}. 
In this analysis, we fix the redshift at $z=7.1899$ and include all the FIR and mm photometry observed with Herschel, SCUBA-2, and NOEMA presented in \cite{Fujimoto+2022}. 
For the stellar component, we adopt the single stellar population models of \cite{BC03} (\texttt{bc03} module), assuming a \cite{Chabrier2003_IMF} initial mass function (IMF). 
The SFH is modeled using a delayed-$\tau$ function. We also include a burst component in the model whose SFH can be modeled as an exponential profile. 
The mass fraction of the late starburst population is allowed to vary between 0 and 0.5 in step of 0.1. 
Following procedures of \cite{Zhuang+2023_AGN} and \cite{Tanaka+2025}, we set an upper limit on the age of the stellar population, requiring it to be less than 95\% of the age of the universe at the galaxy's redshift ($t_{\rm age}\leq 0.95\,t_{\rm H}$). 
We include contributions from ionized gas using the \texttt{nebular} module. To account for the potentially extreme interstellar medium (ISM) conditions in high-redshift galaxies, we allow the ionization parameter ($\log U$) to vary between $-$3 and $-$1 and the gas-phase metallicity to range from 10\% to 100\% of the solar value. 
The dust attenuation is also included in the fitting, following a modified \cite{Calzetti+2000} attenuation law (\texttt{dustatt\_modified\_starburst} module). 
Given the confirmed presence of broad lines in our source, we include an AGN component using the default \texttt{skirtor} module in \texttt{CIGALE} \citep{Stalevski+2016_torus, Boquien+2019_cigale}. 
We fix the viewing angle to a face-on orientation ($i=0^\circ$) in the fitting and let AGN emission vary uniformly from 0.1 to 0.9 in the rest-UV-to-optical band (1000\AA-10000\AA). 

In Figure~\ref{fig3: sed}, we present our best-fit SED with \cigale\ (black), together with the individual components of stellar (blue), nebular (green), AGN (red), and thermal dust emission (yellow) from the stellar component. 
Our best-fit results suggest that the rest-frame UV-to-optical regime is dominated by the AGN continuum, which is consistent with the independent galaxy+AGN composite SED fitting presented in \cite{Fujimoto+2022}.  
In the rest-frame near-infrared (NIR) bands, the stellar emission can be comparable to the AGN emission, which is consistent with our 2D decomposition results. 
In the rest-frame mid-infrared (MIR) regime, the emission is subsequently dominated by thermal emission from the hot dust torus, which is also consistent with the absence of the residual component in the F1800W map after subtracting the point source component (Figure~\ref{fig2: galfits}). 
At the longest wavelengths in the rest-frame far-infrared band, the SED is primarily shaped by the thermal emission originating from cold dust associated with the star-forming activity of the host galaxy. 
The SED fitting yields a host stellar mass of $\log(M_*/M_\odot) = 10.5\pm0.4$, and a star formation rate of $\mathrm{SFR}=330\pm97\,M_\odot\,\rm yr^{-1}$.

We also estimated the SFR using the Paschen-$\alpha$ (Pa$\alpha$) emission line with the MIRI spectrum. 
The spectrum was extracted from a circular aperture with a 0\farcs5 radius to capture the central emission region. 
We modeled the emission feature using a single Gaussian profile superimposed on a polynomial continuum (Figure~\ref{fig4:miri}). 
Notably, the line width of the Pa$\alpha$ (${\rm FWHM}=247\pm61\,\rm km\,s^{-1}$) is comparable to the \cii\ emission line from NOEMA observation \citep[${\rm FWHM}=280\pm40\,\rm km\,s^{-1}$, Tab~\ref{tab1:source}, ][]{Fujimoto+2022}, and indicates that a broad component is not required to describe the observed Pa$\alpha$ line profile, suggesting the emission is dominated by star formation. 
We measured an integrated Pa$\alpha$ flux of $88\pm21\times 10^{-19}\rm erg\,s^{-1}\,cm^{-2}$, which implies an SFR of $250\pm88\,M_\odot\,\rm yr^{-1}$. 
This spectroscopic estimate aligns with the SFR derived from our broadband SED fitting within the uncertainties, confirming that the host galaxy of \targ\ is undergoing vigorous star formation.

However, we note that the current SED fit yields a lower $L_{\rm IR}$ and SFR$_{\rm IR}$ compared to previous estimates \citep[$L_{\rm IR}=(1.6\pm0.3)\times 10^{13}\,L_\odot$; ][]{Fujimoto+2022}, possibly due to an underprediction of the SCUBA 850~$\mu$m flux (Figure~\ref{fig3: sed}). 
Previous work derived the SFR$_{\rm IR}$ using a modified blackbody fit applied only to the FIR photometry, a method that bypasses the complexities of enforcing energy balance against a quasar-dominated UV/optical continuum. 
That approach successfully reproduced the 850 $\mu$m excess, resulting in a higher inferred $L_{\rm IR}$. 
It is worth noting that discrepancies where ${\rm SFR}_{\rm IR} > {\rm SFR}_{\rm Pa\alpha}$ have been observed in high-$z$ dusty galaxies \citep{Alvarez-Marquez+2023_dsfg, Bik+2024_dusty}. 
This excess suggests either that heavy obscuration is attenuating even the Paschen lines, or that we are witnessing a difference in star formation timescales, where Pa$\alpha$ traces instantaneous activity ($\sim$10 Myr) while the FIR emission integrates heating over a longer period ($\sim$100 Myr).

\subsection{$M_{\rm BH}-M_*$ relation}
\label{sec4.3: relation}


The observed tight scaling relations between $M_{\rm BH}$ and host galaxy bulge properties, such as the $M_{\rm BH}-\sigma$ and $M_{\rm BH}-M_{\rm bulge}$ correlations \citep{Magorrian+1998, Ferrarese+2000, Gebhardt+2000, Kormendy+2013}, indicate a fundamental co-evolutionary pathway linking these components \citep{Madau+2014}. 
A similar, albeit more dispersed, correlation is also established between $M_{\rm BH}$ and the total stellar mass ($M_*$) \citep[e.g., ][]{Reines+2015}. 
This coupled growth is hypothesized to be particularly critical during the earlier phases of galaxy and BH formation. 
However, contrasting with these well-established local benchmarks, AGNs observed at high redshifts ($4.5<z<7$) present a significant deviation. 
These high-$z$ sources systematically populate the region above the extrapolated local $M_{\rm BH}-M_*$ relationship and are characterized by significant scatters \citep[e.g., ][]{Harikane+2023, Maiolino+2024}.

A prevailing explanation for this discrepancy has been observational selection bias, where flux-limited surveys are inherently biased toward detecting only the most luminous and massive BHs whose emission outshines that of their host galaxy \citep[e.g., ][]{Shen+2015, LiJ+2021, LiJ+2025b, Tanaka+2025, Silverman+2025}.  
However, recent deep JWST surveys focusing on low-luminosity AGNs at high redshift have argued that this selection effect is insufficient to fully account for the large observed offset, implying a genuine physical difference in the galaxy-BH connection at high redshift \citep[e.g., ][]{Pacucci+2023, Maiolino+2024, Stone+2024_QSO}. 

\begin{figure*}
    \centering
    \includegraphics[width=0.7\linewidth]{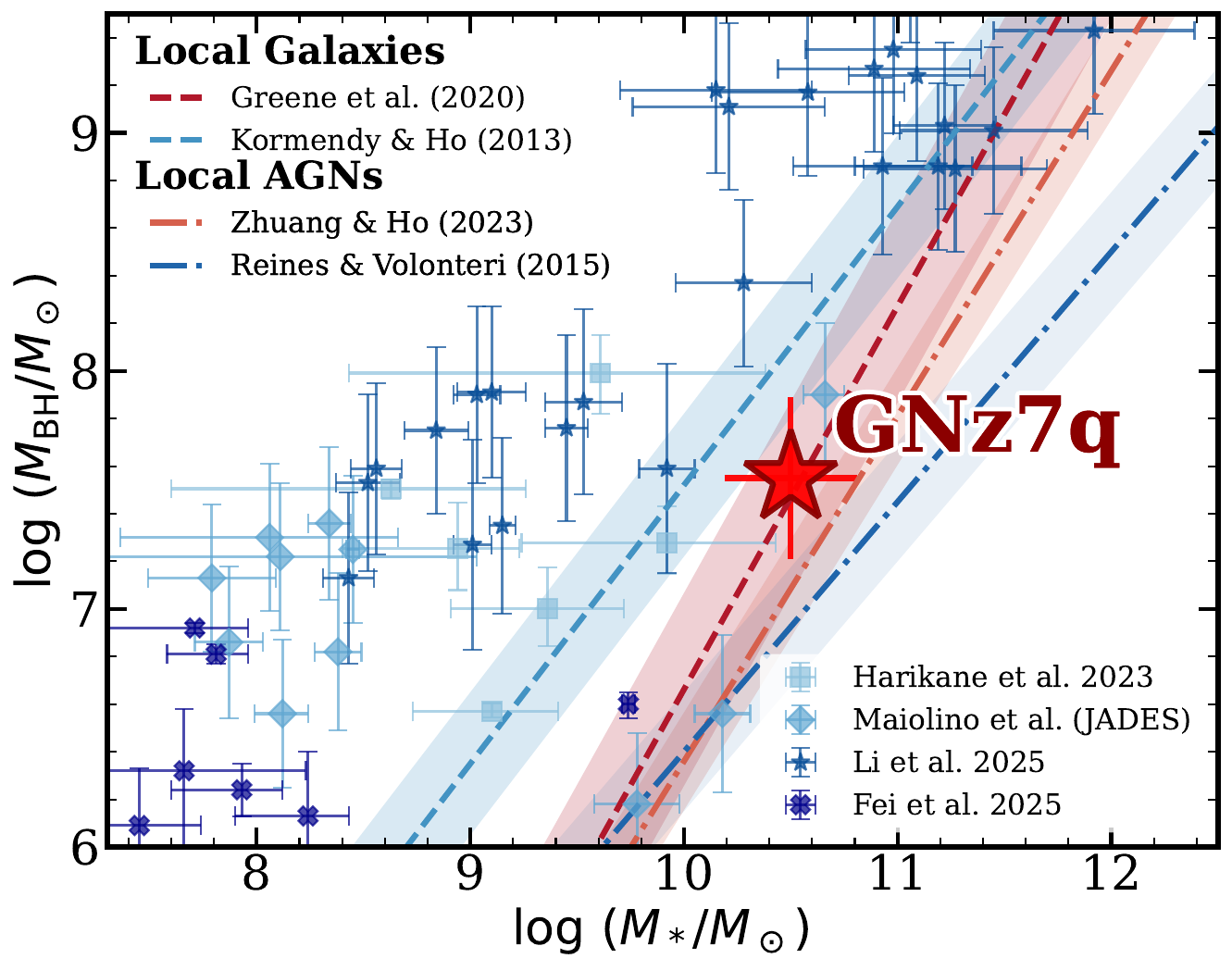}
    \caption{
    BH mass vs. stellar mass for \targ\ and other high-$z$ AGNs, including the JWST discovered faint AGNs at $z\gtrsim 4$ \citep{Harikane+2023, Maiolino+2024, Fei+2025} and luminous quasars at $z\sim 6$ \citep{Fan+2023, Yang+2023_qso, Li+2025_galfits}. The dashed and dotted lines represent the empirical relationships for the local galaxies \citep{Kormendy+2013, Greene+2020} and the local AGNs \citep{Reines+2015, Zhuang+2023_AGN}, respectively. \targ\ is shown as the red star. }
    \label{fig4: rel}
\end{figure*}

Based on the derived $M_{\rm BH}$ and $M_*$, we calculate a mass ratio of $M_{\rm BH}/M_* = 0.1\%$, and the $M_{\rm BH}-M_*$ relation for \targ\ as well as other high-$z$ AGNs are shown in Figure~\ref{fig4: rel}. 
Our analysis demonstrates that the derived $M_{\rm BH}$ and $M_*$ values align with the established $M_{\rm BH}-M_*$ scaling relationship defined in the local Universe \citep[e.g., ][]{Kormendy+2013, Greene+2020}. 
This consistency suggests that some AGN-galaxy systems can achieve a state of co-evolutionary maturity as early as $z \sim 7$ \citep{Silverman+2025, LiJ+2025b, Onoue+2025_shellq}. 


\subsection{Potential Evolutionary Pathway}
\label{sec4.4: evolution}

\begin{figure*}
    \centering
    \includegraphics[width=0.495\linewidth]{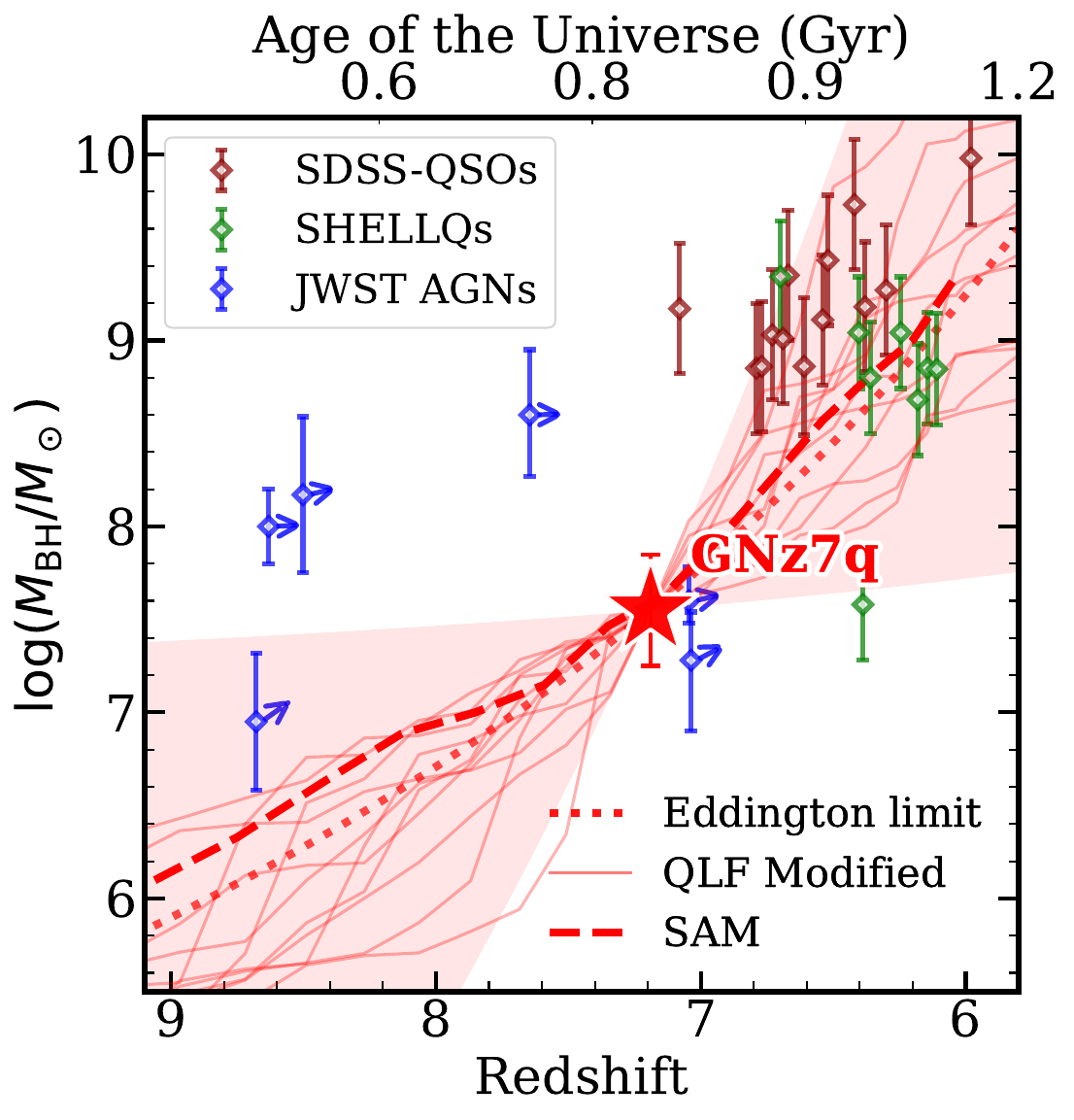}
    \includegraphics[width=0.495\linewidth]{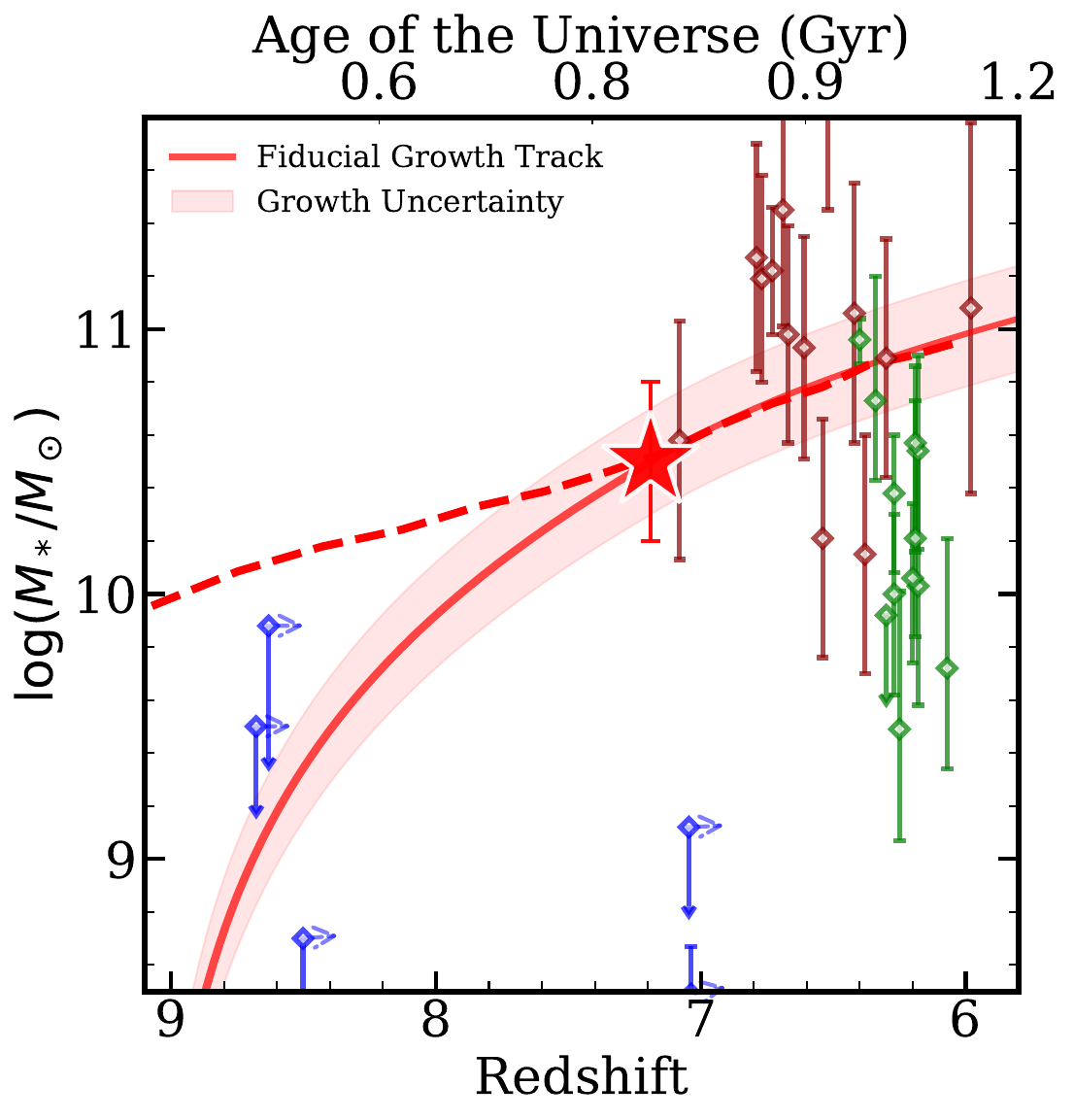}
    \caption{
    Evolutionary pathways of \targ\ and comparative high-$z$ samples.
    Left: $M_{\rm BH}$ vs. redshift. The dark-red, green, and blue diamonds denote luminous quasars \citep{Fan+2023}, low-luminosity quasars selected from Subaru HSC \citep[e.g., ][]{Matsuoka+2018_SHELLQ, Onoue+2025_shellq} and JWST faint AGNs \citep[e.g., ][]{Kokorev+2023, Akins+2025, Furtak+2024_len_LRD, Larson+2023, Taylor+2025_HzLRD}, respectively. 
    Theoretical evolutionary tracks are adopted from \cite{Li+2024_QLF} (light-red solid line) and the DELPHI semi-analytic model \citep[red dashed line; ][]{Dayal+2019_DELPHI}. 
    The red dotted line indicates a baseline of assuming continuous Eddington-limited accretion ($\lambda_{\rm Edd}=1$), with the red-shaded region denoting the possible region with $0.1<\lambda_{\rm Edd}<2.7$.
    Right: $M_*$ vs. redshift for our targets and comparison sample. 
    The red shaded region shows the mass growth for \targ\ host galaxy based on the SFH derived from the SED fitting (Table~\ref{tab1:source}), and the red shaded line represents the growth track of $M_*$ from DELPHI semi-analytic model. 
    }
    \label{fig5: evo-track}
\end{figure*}

The suite of derived physical parameters, including $M_{\rm BH}$ and $\lambda_{\rm Edd}$, alongside the $M_*$ and SFR, affords crucial constraints on the system's evolutionary trajectory. 
When compared with other high-$z$ AGNs, specifically the luminous ground-based quasars and the faint AGNs detected by JWST, this source occupies an intermediate locus in the census of $M_{\rm BH}$ and $M_*$, i.e., $M_{\rm BH}$ and $M_*$ of \targ\ are greater than those of faint AGNs while smaller than luminous quasars.
More importantly, \targ\ is characterized by rapid dual growth in the manner of both BH and host galaxy, driven by high Eddington ratios and substantial star formation activity. 
This unique feature suggests that \targ\ represents a distinct phase on the BH-galaxy co-evolutionary track.

To investigate the potential evolutionary trajectory of \targ, we model the evolution of its $M_{\rm BH}$ and $M_*$ as a function of cosmic time. 
We investigate the evolutionary tracks of black hole mass using two physically motivated theoretical frameworks and one simplified baseline. 
First, we employ the QLF-regulated growth model presented in \citep{Li+2024_QLF}, which reconstructs BH growth via episodic accretion calibrated against the observed luminosity functions at high redshift. 
Second, we utilize the semi-analytic DELPHI framework \citep{Dayal+2019_DELPHI}. 
We contrast these realistic scenarios with a simple model assuming continuous accretion at a constant Eddington ratio of unity ($\lambda_{\rm Edd}=1$). 
Regarding the host galaxy assembly, we adopt a simplified assumption of growth via a continuous, constant SFR. 
While we recognize that reconstructing the full Star Formation History (SFH) offers a more realistic perspective \citep[e.g., ][]{Onoue+2025_shellq}, we opt for the constant SFR approximation to mitigate the substantial uncertainties currently inherent in constraining detailed SFHs at these redshifts. 
For comparison, we also calculate the corresponding evolutionary paths for other AGN populations, including highly luminous ground-based quasars and faint AGNs, following the methodology detailed in \cite{Li+2025_galfits}. 
The $M_{\rm BH}$ and accretion rates are adopted from previous studies \citep[e.g., ][]{Kokorev+2023, Furtak+2024_len_LRD, Tripodi+2024, Larson+2023}. 
The SFRs for these high-$z$ AGN host galaxies are determined by combining literature values derived from the rest-UV luminosities with recent average constraints obtained from ALMA observations \citep[e.g.,][]{Casey+2025}.

Figure~\ref{fig5: evo-track} presents the modeled evolutionary tracks for \targ\ and the faint AGN comparison sample at $z>7$. 
Assuming a realistic SFR and the $\dot{M}_{\rm BH}$, the modeled evolution for \targ\ demonstrates a potential increase in $M_{\rm BH}$ by approximately $1$ dex (with a maximum of $\sim 2$ dex) and an increase in $M_*$ by $\sim 0.5$ dex. 
This modeled growth path converges with the locus of luminous quasars observed in large surveys (e.g., SDSS, HSC) at $z\gtrsim 6$, suggesting that \targ\ is a likely progenitor for these massive systems. 

In contrast, we find that the faint AGNs recently discovered by JWST at similar or higher redshifts are challenging to be the direct progenitors of the SDSS-like quasars. 
For the BH mass, though their large masses make them possible for them to become SMBH at $z\sim 6$ \citep{Larson+2023, Taylor+2025_AGN}, the BH mass measurements still have large uncertainties due to distinct models \citep{Inayoshi2025_LRD, Inayoshi+2025_LRD, Naidu+2025_BH*, Greene+2025_lrd_lbol}. 
Additionally, the relatively low SFR as well as the barely-detected host galaxies, in particular for LRD, make their host galaxies hardly evolve to the quasar host galaxies, which are usually massive \citep[e.g., $M_*\gtrsim 10^{10}\,M_\odot$][]{Li+2025_galfits, Ding+2023_nat, Ding+2025_shellq, Onoue+2025_shellq, Fei+2025}, starbursts \citep[e.g., ${\rm SFR}\sim 500-1000\,M_\odot\,\rm yr^{-1}$; ][]{Wang+2013_qso, Decarli+2018_qso, Venemans+2020_qso, Wang+2024_qso}. 
Indeed, their $M_*$ shows limited growth in a short timescale. 
This is attributable to the relatively low inferred SFRs, which is further supported by the non-detection of far-infrared dust emission in these targets, particularly those exhibiting red continua \citep{Akins+2025, Casey+2025}. 

Beyond the comparison of $M_{\rm BH}$ and $M_*$, the fundamental physical distinctions between \targ\ and the LRDs can be inferred from several other aspects. 
First, the nature of the obscuration differs significantly. 
\targ\ exhibits substantial dust reddening consistent with a classic obscured quasar, while LRDs are characterized by a distinct lack of dust \citep[e.g., ][]{Naidu+2025_BH*}. 
Second, the optical \feii\ emission in \targ\ presents as a strong, broad complex, in sharp contrast to the narrow \feii\ features typically observed in LRD spectra \citep[e.g., ][]{Torralba+2025_LRD_FeII, D'Eugenio+2025_LRD_FeII}. 
Finally, we find no evidence in \targ\ for the high column density of neutral gas that is always seen in LRDs \citep{Matthee+2024_LRD, Inayoshi+2025_LRD}, which manifests as significant Balmer breaks or Balmer absorption lines. 
The absence of these features suggests that \targ\ is distinct from the LRD population. 
Instead, these features reinforce the scenario that \targ\ represents a progenitor to the standard population of luminous blue quasars.

Overall, the derived properties and the modeled evolutionary trajectory of \targ\ position it as a plausible direct progenitor of the luminous quasars so far discovered by ground-based facilities during a decades-long search, while the distinct evolutionary pathway of the faint JWST AGNs suggests they may constitute an intrinsically separate population in the high-redshift universe \citep{Inayoshi+2025_lrd_review}. 
A key question then is: How could a system like \targ\ have escaped from the ``stellar-deficient'' earlier stage that may be a legacy of the initial seeding process? 
This possible bifurcation in early growth trajectories could reflect differences in the large-scale cosmic environment, to be assessed in future work. 

\section{Conclusion}
\label{sec5}
In this work, we presented a comprehensive multi-wavelength characterization of \targ, a red quasar at $z=7.1899$, utilizing the full suite of JWST instrumentation (NIRCam, NIRSpec, and MIRI) and archival data. 
This analysis provides novel insights into the rapid black hole growth and the co-evolutionary scenario of galaxy-BH systems in the early Universe. 

Through a detailed analysis of the NIRSpec spectroscopy, we derived the following properties for the central AGN:
\begin{enumerate}
    \item The spectrum exhibits broad Balmer emission lines (\ha, \hb, \hg) and strong \feii\ lines. Kinematic decomposition of the broad lines reveals a double-Gaussian profile, comprising a narrower ($\rm FWHM\approx 2200\,km\,s^{-1}$), systemic component and a significantly broader ($\rm FWHM\approx 11000\,km\,s^{-1}$), highly redshifted ($\Delta v\approx 3600\,\rm km\,s^{-1}$) component.
    \item The Balmer decrement of the narrow emission lines (\ha/\hb\ ratio) and the continuum SED slope suggest a relatively low level of dust attenuation within the broad-line region. Assuming a standard extinction law, we derive an extinction of $A_V=0.21_{-0.09}^{+0.08}$. 
    \item Applying local virial scaling relations yields a BH mass of $\log(M_{\rm BH}/M_\odot) = 7.55\pm0.34$ and a super-Eddington accretion ratio of $\lambda_{\rm Edd}= 2.7\pm0.4$. These properties identify \targ\ as a rapidly growing massive BH in the Epoch of Reionization, exhibiting one of the highest known accretion rates at this redshift.
    \item This high Eddington accretion rate is further supported by the detection of the broad neutral \oi\ emission line, which has been observed in several high-$z$ faint AGNs discovered by JWST. 
    \item The observed X-ray bolometric correction ($k_{\rm bol}$) presents a significant deviation ($>5\sigma$) from the empirical relations, implying a suppression of the X-ray emission in \targ. This depression is likely attributable to the intrinsic effects of a high accretion rate, heavy obscuration by high column density material along the line of sight, or the interplay of both mechanisms.
\end{enumerate}

Utilizing NIRCam and MIRI imaging alongside MIRI spectroscopy, the host galaxy of this system is characterized as:
\begin{enumerate}
\setcounter{enumi}{4}
    \item Morphological decomposition robustly resolves the host galaxy's stellar continuum in the rest-frame optical and near-infrared bands, revealing an effect radius of $R_e=1.01\,\rm kpc$ and a S\'ersic index of $n=1.9$. In the rest-frame mid-infrared, the stellar contribution declines, indicating the transition to a regime dominated by thermal emission from the AGN dust torus. Interestingly, high-resolution rest-frame UV imaging reveals a clumpy morphology characterized by two distinct structures. 
    \item The SED fitting yields a decomposition consistent with standard system components. The rest-frame UV is driven by the AGN, while the host stellar emission becomes important in the rest-optical/NIR band. At mid-IR band, the SED is dominated by the hot torus emission. The thermal emission from the dust in the host galaxy dominates the far-IR and sub-mm emission. 
    \item The far-infrared SED implies a vigorous star formation rate of ${\rm SFR_{FIR}}=330\pm97\,M_\odot\,\rm yr^{-1}$, confirming the host as a dusty starburst. This value is consistent, within uncertainties, with the independent estimate derived from the Pa$\alpha$ emission line (${\rm SFR_{Pa\alpha}}=250\pm88\,M_\odot\,\rm yr^{-1}$).
    \item The SED fitting yields a substantial dust attenuation for the host galaxy ($A_{V,\rm host}=3.1\pm0.8$). This value significantly exceeds the attenuation derived for the central AGN component, pointing to distinct dust distributions obscuring the host galaxy and the central engine, respectively.
\end{enumerate}

The derived properties of the BH and host galaxy position \targ\ in a unique phase of the BH-galaxy co-evolution scenario: 
\begin{enumerate}
\setcounter{enumi}{8}
    \item In contrast to many JWST-identified faint AGNs and high-$z$ quasars discovered from all-sky surveys that appear over-massive relative to their hosts, \targ\ exhibits a mass ratio of $M_{\rm BH}/M_* \approx 0.1\%$, placing it consistently on the local $M_{\rm BH}-M_*$ scaling relation. This alignment indicates that the BH-galaxy relation can be achieved as early as $z=7.1899$. 
    \item In conclusion, \targ\ offers a rare glimpse into the progenitor phase of luminous quasars, illustrating a distinct stage of early co-evolution before the onset of the unobscured, SDSS-like quasar regime. 
\end{enumerate}

The modeled evolutionary trajectory of \targ\ identifies it as a dust-obscured, super-Eddington progenitor likely representing an early assembly phase of luminous $z \sim 6$ quasars.
Consequently, we anticipate that future deep surveys targeting higher redshifts ($z \gtrsim 8-10$) will unveil a population of progenitor systems that mirror the properties of \targ. 
Crucially, our growth models suggest these earlier sources will not exhibit the extreme `overmassive' BH mass ratios seen in most faint AGN populations; instead, they should populate the extension of the local $M_{\rm BH}-M_*$ scaling relation. 
This prediction implies that the physical coupling between BH accretion and host galaxy assembly is likely established within the first few hundred million years of the Universe. 

\textit{Facility:} JWST, NOEMA

\section*{Acknowledgements}
We are grateful to Jinyi Yang and Feige Wang for their insightful discussions about the broad line fitting. 
We are grateful to Zhiwei Pan and Yuming Fu for their discussions about the BH mass estimator. 

K.I. acknowledges support from the National Natural Science Foundation of China (12573015, W2532003), the Beijing Natural Science Foundation (IS25003), and the China Manned Space Program (CMS-CSST-2025-A09).
LCH was supported by the National Science Foundation of China (12233001) and the China Manned Space Program (CMS-CSST-2025-A09). 
PGP-G acknowledges support from grant PID2022-139567NB-I00 funded by Spanish Ministerio de Ciencia, Innovaci\'on y Universidades MCIU/AEI/10.13039/501100011033, FEDER {\it Una manera de hacer Europa}. 
J.A.-M. acknowledges support by grants PID2024-158856NA-I00 \& PIB2021-127718NB-I00 from the Spanish Ministry of Science and Innovation/State Agency of Research MCIN/AEI/10.13039/501100011033 and by “ERDF A way of making Europe”.

This work is based on observations made with the NASA/ESA/CSA James Webb Space Telescope. The data were obtained from the Mikulski Archive for Space Telescopes at the Space Telescope Science Institute, which is operated by the Association of Universities for Research in Astronomy, Inc., under NASA contract NAS 5-03127 for JWST. 
The specific observations analyzed can be accessed via doi: \href{https://archive.stsci.edu/doi/resolve/resolve.html?doi=10.17909/q19z-a348}{10.17909/q19z-a348} and \href{https://archive.stsci.edu/doi/resolve/resolve.html?doi=10.17909/0akt-xx69}{10.17909/0akt-xx69}
These observations are associated with programs \#4762 and \#5407. 

\textit{Facility:} JWST

\textit{Software:} \texttt{LMFIT} \citep{Newville+2014_lmfit} \texttt{grizili} \citep{Brammer+2021_grizili} \texttt{msaexp} \citep{Brammer2023}


\bibliography{sample631}{}
\bibliographystyle{aasjournal}

\appendix 
\counterwithin{figure}{section}

\section{Evaluation of Additional Spectral Components} 
\label{sec3.2: eva}
The inherent complexity and multi-component nature of quasar spectra can make the fitting results highly dependent on the adopted model. 
We therefore performed a model comparison, evaluating a suite of models with different combinations of components and parameter constraints. 
To identify the optimal description, we employed the Bayesian Information Criterion (BIC):
\begin{align}
    \mathrm{BIC} \equiv \chi^2+k\ln n,
\end{align}
where $k$ is the number of free parameters and $n$ is the number of data points. 
The model that yielded the minimum BIC value, representing the best compromise between goodness-of-fit ($\chi^2$) and model simplicity (penalized by $k$), was adopted as our final best-fit.

\begin{figure}
    \centering
    \includegraphics[width=\linewidth]{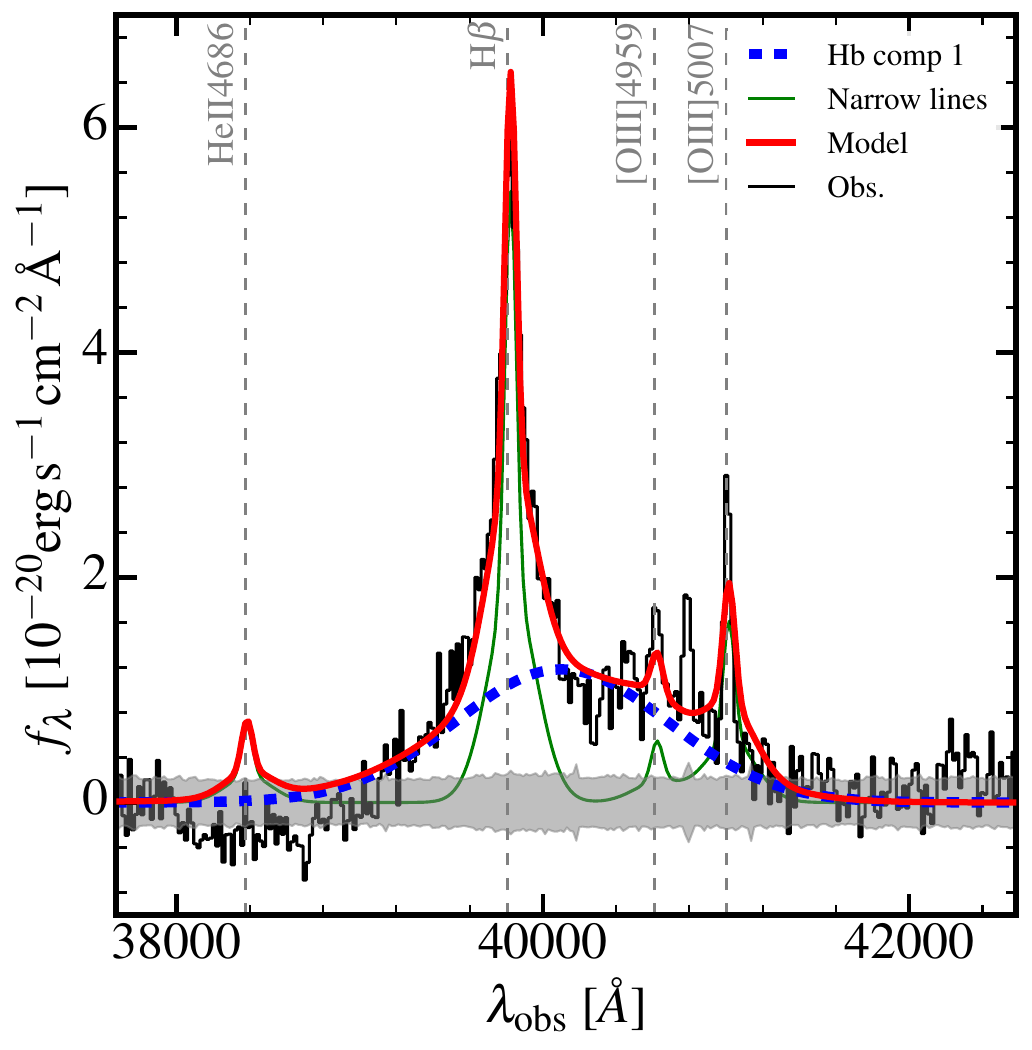}
    \caption{
    Spectral fit of the H$\beta$+\oiii\ complex adopting a single-component model for the broad emission line. 
    The continuum subtracted spectrum is shown as black solid line, and the best-fit model is shown as red thick line. 
    The narrow line model and the broad line model are shown as a green line and a blue dashed line, respectively. 
    }
    \label{figA1}
\end{figure}

We firstly performed a model comparison to determine the optimal number of Gaussian components required to characterize the broad Balmer lines. 
A simple, single-component model is strongly disfavored, yielding a $2,000\,\rm km\,s^{-1}$ redshift relative to the systemic velocity and a significantly poorer fit ($\Delta \rm BIC=116$) compared to the model in Section~\ref{sec3.1} (Figure~\ref{figA1}). 
Conversely, models incorporating three or more components provided no statistically significant improvement. 
We thus conclude that our two-component model provides the most parsimonious and statistically robust description of this source's broad line region. 
This redshifted broad \hb\ component has been observed in other sources at different redshifts, in particular for sources with high Eddington accretion rates \citep{Marziani+2009}, whose origin is still under debate, while all potential explanations regard it as a non-virial component within the BLR. 
Therefore, we estimate the BH and AGN properties only using the nBLR. 

We also consider the possibility that the Iron emission lines have the same broadening and shifting compared to the \hb\ emission line\citep{Sulentic+2012_FeII}. 
In this consideration, we fixed the line width of the \feii\ lines to that of the broad \hb\ component (the extremely broad component is not included). 
However, it failed to reproduce prominent features in the Iron pseudo-continuum, particularly the emission spikes in the rest-frame 5100--5500$\,\AA$. 
More importantly, this model is disfavored by a significantly larger BIC ($\Delta \rm BIC=110$). 
Therefore, we conclude that the iron-emitting region is kinematically decoupled from the broad line region in this source. 
Our best-fit model, which allowed the \feii\ width to be a free parameter, converged on a significantly narrower profile, with $\rm FWHM_{Fe} / FWHM_{H\beta} \approx 46\%$. 
This model comparison confirms that our fiducial setup (Section~\ref{sec3.1}) is preferred.

\begin{figure*}
    \centering
    \includegraphics[width=0.8\linewidth]{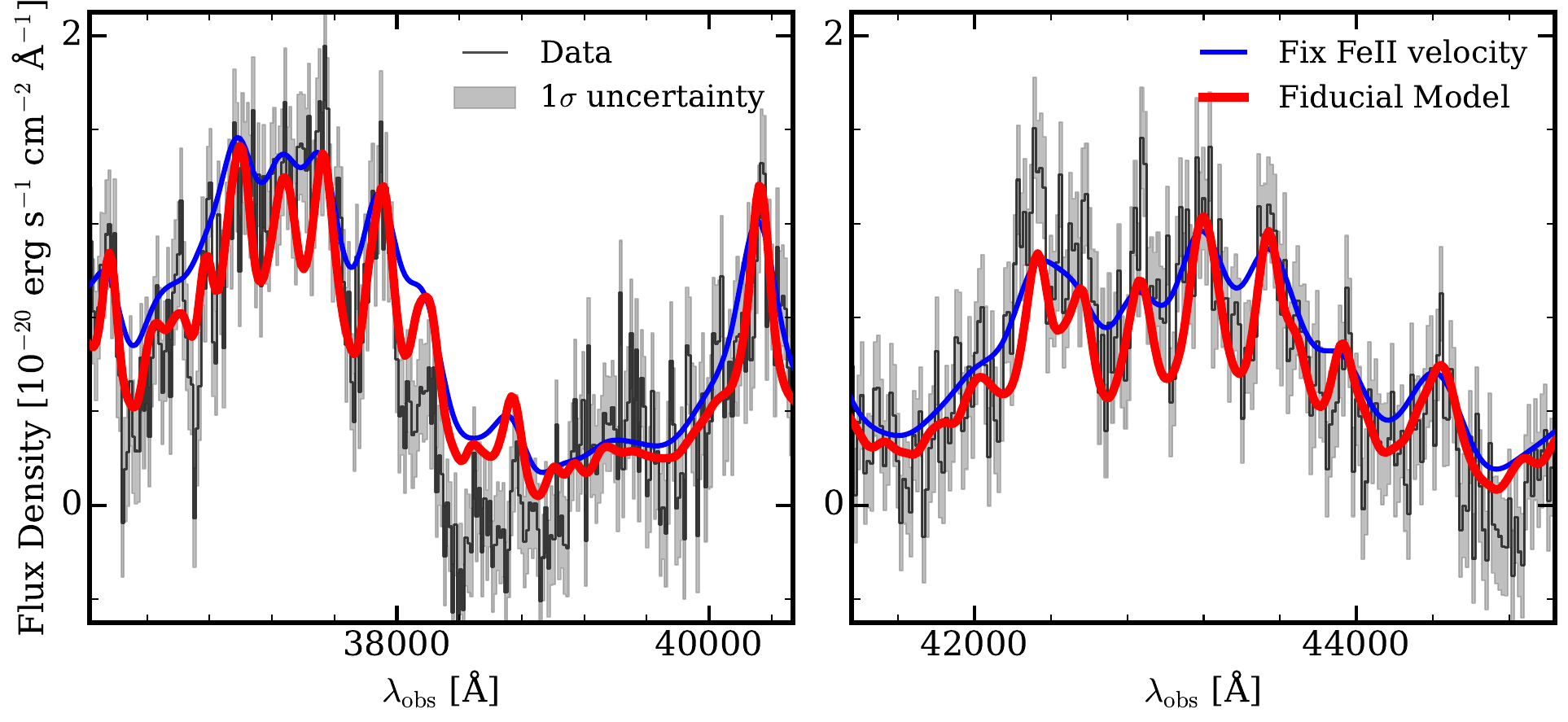}
    \caption{
    Comparison of different kinematic models for Iron. 
    The observed spectrum is shown in black for the rest-frame 4400–4950 \AA and 5050–5500 \AA regions. 
    The blue line indicates the Iron pseudo-continuum with the line width constrained to match the broad H$\beta$ profile, while the red line represents our fiducial model.
    }
    \label{figA2}
\end{figure*}

\section{MIRI LRS spectrum}
\label{app2: MIRI LRS}

\begin{figure*}
    \centering
    \includegraphics[width=0.7\linewidth]{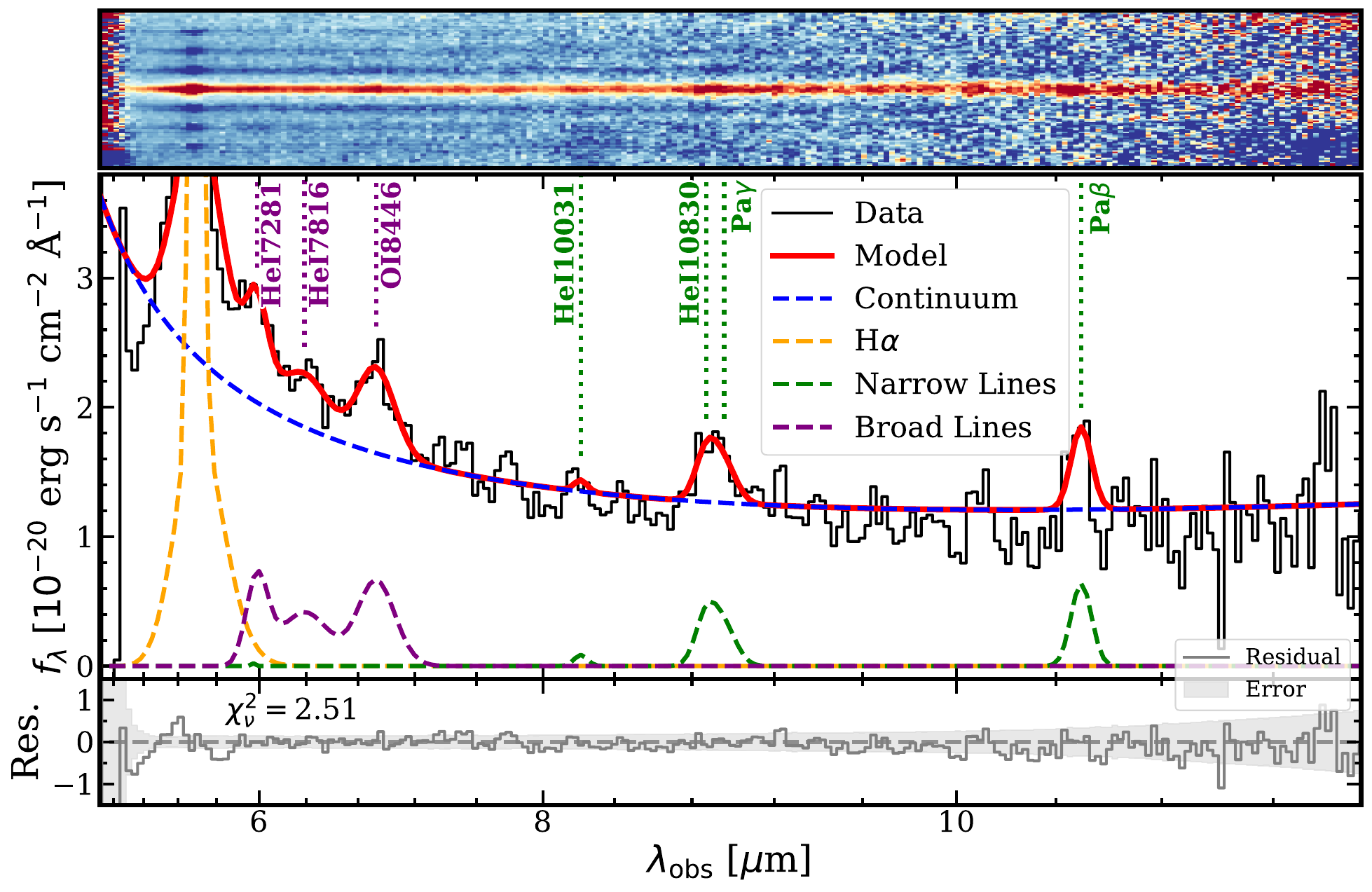}
    \caption{The observed rest-frame optical-to-NIR spectrum, generated from the MIRI LRS observation. 
    \textit{Top:} 2D spectrum from the MIRI LRS observation.
    \textit{Middle:} 1D spectrum extracted from the 2D spectrum and the corresponding model. 
    Black and red solid lines represent the observed spectrum and the resultant best-fit model, respectively. 
    Component modeling includes a power-law continuum (blue dashed line), \ha\ emission (orange dashed line), and decoupled narrow (green) and broad (purple) emission line profiles. 
    The corresponding emission lines are labeled with different colors (purple for broad lines and green for narrow lines). 
    Note that we do not consider the instrumental broadening during the fitting, the broadening of those rest-frame NIR lines can be caused by the instrument. 
    \textit{Bottom:} The residual between the data and the model. The 1$\sigma$ error is shown as the gray shaded region. }
    \label{fig1.1: LRS}
\end{figure*}

Figure~\ref{fig1.1: LRS} illustrates the two-dimensional and extracted one-dimensional MIRI LRS spectra, accompanied by the best-fit model and corresponding residuals. 
Due to the limited spectral resolution inherent to the MIRI LRS instrument, the line widths for the Near-Infrared (NIR) features were not constrained in this analysis.

\end{document}